\documentclass[11pt]{emulateapj}
\usepackage{natbib}
\usepackage{graphicx}
\usepackage{amsmath}
\usepackage{amssymb}
\def\Wm2{W/m$^2$}
\def\Wpm2sr{Wm$^{-2}sr^{-1}$}
\def\deg{$^\circ$ }
\def\degx{$^\circ$}

\def\mum{$\mu$m }
\def\mumx{$\mu$m}

\begin{document}
\title{Post-equinox dynamics and polar cloud structure on
    Uranus\footnotemark[\dag]} 
\author{L.~A. Sromovsky\altaffilmark{1},
  P.~M. Fry\altaffilmark{1},
  H.~B. Hammel\altaffilmark{2}$^,$\altaffilmark{3}, I. de Pater\altaffilmark{4},
  and K.~A. Rages\altaffilmark{5}}
\altaffiltext{1}{University of Wisconsin - Madison, Madison WI 53706}
\altaffiltext{2}{AURA, 1212 New York Ave. NW, Suite 450, Washington, DC 20005, USA}
\altaffiltext{3}{Space Science Institute, Boulder, CO 80303, USA}
\altaffiltext{4}{University of California, Berkeley, CA 94720, USA}
\altaffiltext{5}{SETI Institute, Mountain View, CA 94043, USA}
\altaffiltext{\dag}{Based in part on observations with the NASA/ESA Hubble Space
Telescope obtained at the Space Telescope Science Institute, which is
operated by the Association of Universities for Research in Astronomy,
Incorporated under NASA Contract NAS5-26555.} 

\slugcomment{Journal reference: Icarus 220 (2012) 694-712.}

\begin{abstract}
Post equinox imaging of Uranus by HST, Keck, and Gemini telescopes
has enabled new measurements of winds over previously sampled
latitudes as well as measurements at high northern latitudes that have
recently come into better view. These new observations also used
techniques to greatly improve signal to noise ratios, making possible
the detection and tracking of more subtle cloud features. The 250 m/s
prograde jet peaking near 60\degx N was confirmed and more accurately
characterized.  Several long-lived cloud features have also been
tracked. The winds pole-ward of 60\degx N are consistent with solid
body rotation at a westward (prograde) rate of 4.3\degx/h with respect
to Uranus' interior.  When combined with 2007 and other recent
measurements, it is clear that a small but well-resolved asymmetry
exists in the zonal profile at middle latitudes, peaking at 35\degx,
where southern winds are 20 m/s more westward than corresponding
northern winds. High S/N Keck II imaging of the north polar region of
Uranus reveals a transition from streaky bands below 60\deg N to a
region from 60\deg to nearly the north pole, where widely distributed
small bright spots, resembling cumulus cloud fields, with several isolated
dark spots, are the dominant style of cloud features.  This presents a
stark contrast to 2003 detailed views of the south polar region of
Uranus when no discrete cloud features could be detected in comparable
Keck II near-IR images.  The pressure levels of discrete clouds
estimated from spatial modulations in H and Hcont images indicate that
the polar cloud features are generally in the 1.3 to 2-3 bar range, as
are equatorial and several mid-latitude features. Several of the
brighter mid latitude features are found above the 1.2-bar level of
methane condensation.

\end{abstract}
\keywords{Uranus, Uranus Atmosphere;  Atmospheres, dynamics}

\maketitle
\shortauthors{Sromovsky et al.} 
\shorttitle{Post-equinox dynamics and polar cloud structure on Uranus}

\section{Introduction}

The last update to the wind profile of Uranus was assembled by
\cite{Sro2009eqdyn} from intensive observations carried out near the
time of the 2007 equinox of Uranus. These results were compared with
past results from 1986 Voyager observations \citep{SmithBA1986}, 1997
NICMOS observations \citep{Kark1998Sci}, 1997-2000 HST observations
\citep{Hammel2001Icar}, a remeasurement of 2003 Keck results
\citep{Hammel2005winds,Sro2009eqdyn}, Keck results from 2003 and 2004
\citep{Sro2005dyn}, 2005 Keck results \citep{Sro2007bright}, and 2006
Keck results \citep{Hammel2009Icar}. Taken together, these data sets
provided evidence for a small asymmetry in the zonal wind profile, but
no substantial evidence for any systematic long-term temporal
variation that might be associated with seasonal variations. The 2007
observations finally were able to detect cloud features up to 70\degx
N, and obtained precise wind measurements up to 62\degx N
(planetocentric) latitude, for the first time showing the beginnings
of a northern prograde jet peak with a speed near 250 m/s.

Since 2007 the north polar region of Uranus
has become much better exposed to view as the sub-solar latitude has moved
 to 14\degx N during the 2011 opposition.  This gave us a chance to
better  determine the zonal wind profile in the northern hemisphere and provide a
much better characterization of the prograde jet.
In the following we describe the post equinox observations that we
used to better define the circulation of Uranus, the measurement
results from each data set, new views of the north polar region of
Uranus and the different styles of discrete cloud features located
there, and finally we describe altitude constraints on a subset of
these features.

\begin{table*}\centering
\caption{Telescope/camera characteristics.}
\begin{tabular}{c c  c  c  c c}
\hline\\[-0.1in]
           & Mirror &           &  Pixel & Diff Lim & \\
Telescope  & Diam. &  Camera &    size &    @ Wavelength \\
\hline\\[-0.1in]
HST       &  2.3 m   &   WFC3    &   0.04$''$        &  0.09$''$  @ 0.85 \mum \\
Gemini-North    &  8 m     &   NIRI    &   0.0218$''$    &  0.05$''$  @ 1.6 \mum  \\
Keck II   &  10 m    &   NIRC2-NA &  0.00994$''$     &  0.04$''$  @ 1.6 \mum  \\
Keck II   &  10 m    &   NIRC2-WA &  0.039686$''$    &  0.04$''$  @ 1.6 \mum  \\
\hline\\[-0.1in]
\end{tabular}
\vspace{0.1in}
\parbox{3.8in}{NOTES: Both groundbased telescopes have adaptive
  optics capability, but only Keck II can use Uranus itself as the
  wave front reference. Our Gemini observations had to use a satellite
  of Uranus for the wave front reference.}
\label{Tbl:cameras}
\end{table*}

\section{Observations}

The observations were made with four cameras on three observatories.
The camera characteristics for each observing configuration are given
in Table \ref{Tbl:cameras}, and the imaging observations of Uranus
acquired between 2009 and 2011 are listed in
Table\ \ref{Tbl:tracklist}.  The SNAP observations, acquired in
October 2009 and July and August 2010, do not provide temporal
sampling adequate to measure wind speeds within each SNAP data set.
However, they can provide useful constraints on the motions of long
lived features when combined with other observations.  The November
2009 HST observations do provide temporal sampling adequate to obtain
precise wind measurements, although the number of cloud targets
contained in these images is small because of limited spatial
resolution and inherently low contrast of cloud features at the
observed wavelengths. (Longer wavelengths of higher contrast were not
used because the required WFC3 IR camera has a much coarser pixel
scale of 0.13 arcseconds/pixel.) The HST imaging program was designed
to achieve high S/N ratios to compensate for the inherently low
contrast of the images, which made it possible to obtain 6 wind
vectors with an accuracy of 3-12 m/s, and one with an accuracy of 33
m/s \citep{Fry2012}.  The 3 and 4 June 2010 HST observations provide
complete longitude coverage, but detected only a single discrete cloud
feature near 29\deg N, and did not provide temporal sampling adequate
to determine an accurate wind vector.  The July and August 2010
observations also each provide only a small temporal span and thus can
only be used to track long-lived features if any are observed. There
are two relatively prominent features and one faint feature in the
July 17 images and also three similar features at similar latitudes in
the 5 August images.  However, it is unclear if these are the same
features.

\begin{table*}\centering
\caption{Imaging observations used to track discrete cloud features.}
\begin{tabular}{c c c c c c c l}
\hline\\[-0.1in] 
 Date & Time Range & Telescope/Program & PI & Filters (images) \\  
\hline\\[-0.1in] 
13 Oct. 2009 & 4:36-4:54 & HST/SNAP 11630 & KAR & F845M (12) \\
11 Nov. 2009 & 13:16-22:00 & HST/11573 & LAS & F845M (48)\\
12 Nov. 2009 & 13:14-13:59 & HST/11573 & LAS & F845M (48)\\[0.1in]
3 June 2009 & 9:44-15:14 &  HST/11573 & LAS & 
\parbox{2.5in}{F467M, F658N(2), FQ727N, FQ750N, FQ937N(2),
FQ889N(2), FQ906N(2), FQ924N(2), FQ619N, F547M, F763M, F845M}\\[0.2in]
4 June 2009 & 14:30-15:12 &  HST/11573 & LAS & 
\parbox{2.5in}{FQ619N, F547M, F658N, F763M, F845M, 
FQ924N, FQ906N, FQ889N, FQ937N}\\[0.15in]
17 July 2010 & 3:55-4:12 &  HST/SNAP 11630 & KAR &  F845M (12) \\
5 Aug. 2010 & 1:48-2:54 &  HST/SNAP 11630 & KAR &  F845M (12) \\[0.1in]
2 Nov. 2010 &       &  Gemini-N/2010B-Q-110 & LAS &  \parbox{2.5in}{J(4), H(4), K$'$(8), Hcont(4), CH4L(4)}\\[0.1in]
3 Nov. 2010 &       &  Gemini-N/2010B-Q-110 & LAS &  \parbox{2.5in}{J(4), H(4), K$'$(8), Hcont(4), CH4L(4)}\\[0.1in]
26 July 2011 & 10:51-15:51 & Keck/NIRC2 N128N2 & LAS &  
   \parbox{2.5in}{H(87), Hcont(19), K$'$, CH4\_long}\\[0.02in]
27 July 2011 & 14:02-14:09  & Keck/NIRC2 N128N2 & LAS &  
   \parbox{2.5in}{H(4), Hcont(3)}\\
28 July 2011 & 10:51-15:32 & Keck/NIRC2 N128N2 & IDP & 
 \parbox{2.5in}{H(30), K$'$(7)}\\
23 Oct. 2011 & 5:34-6:29 & Gemini-N/2011B-Q-105 & LAS &  \parbox{2.5in}{H(10), Hcont(10)}\\
25 Oct. 2011 & 5:48-9:01 & Gemini-N/2011B-Q-105 & LAS &  \parbox{2.5in}{H(11), Hcont(19)}\\
26 Oct. 2011 & 5:05-8:58 & Gemini-N/2011B-Q-105 & LAS &  \parbox{2.5in}{H(40), Hcont(72)}\\
\hline\\[-0.1in]
\end{tabular}
\vspace{0.2in}
\parbox{5.8in}{NOTE: Times are UTC. The numbers in the HST filter names correspond to their
 central wavelengths in nanometers. Many of the listed images need to be combined to obtain
adequate S/N ratios.}
\label{Tbl:tracklist}
\end{table*}

\begin{figure}[!htb]\centering
\includegraphics[width=3.2in]{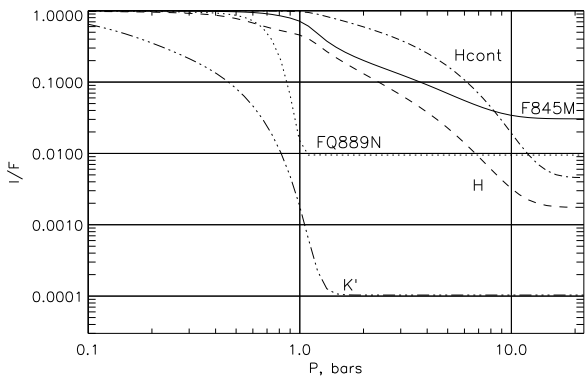}
\caption{I/F for a unit-albedo surface in a clear Uranian atmosphere
  as a function of the pressure at which the surface is placed.  This
  shows how much cloud reflectivity in each filter is attenuated by
  atmospheric absorption above the cloud. The high pressure limits
  correspond to the reflectivity of a clear atmosphere.}
\label{Fig:penprof}
\end{figure}

In Table \ref{Tbl:cameras}, the pixel scale of 0.2138$\pm$0.0005
arcsec/pixel listed for the Gemini NIRI camera was derived by us from
measurements of Uranus and its satellites in comparison with HORIZON
ephemeris positions.  This scale applies when the field lens is used
and differs by 2\% from the standard pixel scale of 0.219
arcseconds/pixel given by the NIRI instrument web page or the value of
0.218 arcseconds/pixel listed under the file header PIXSCALE keyword,
which is only valid without the field lens. We were motivated to
measure the pixel scale by observed deviations from uniform motion as
cloud targets approached the limb of the planet. That problem went
away after the pixel scale was revised. The penetration of these
filters into a clear Uranus atmosphere are shown in
Fig.\ \ref{Fig:penprof}.

\section{Image processing and navigation}

\begin{figure*}[!htb]\centering
\includegraphics[width=5.5in]{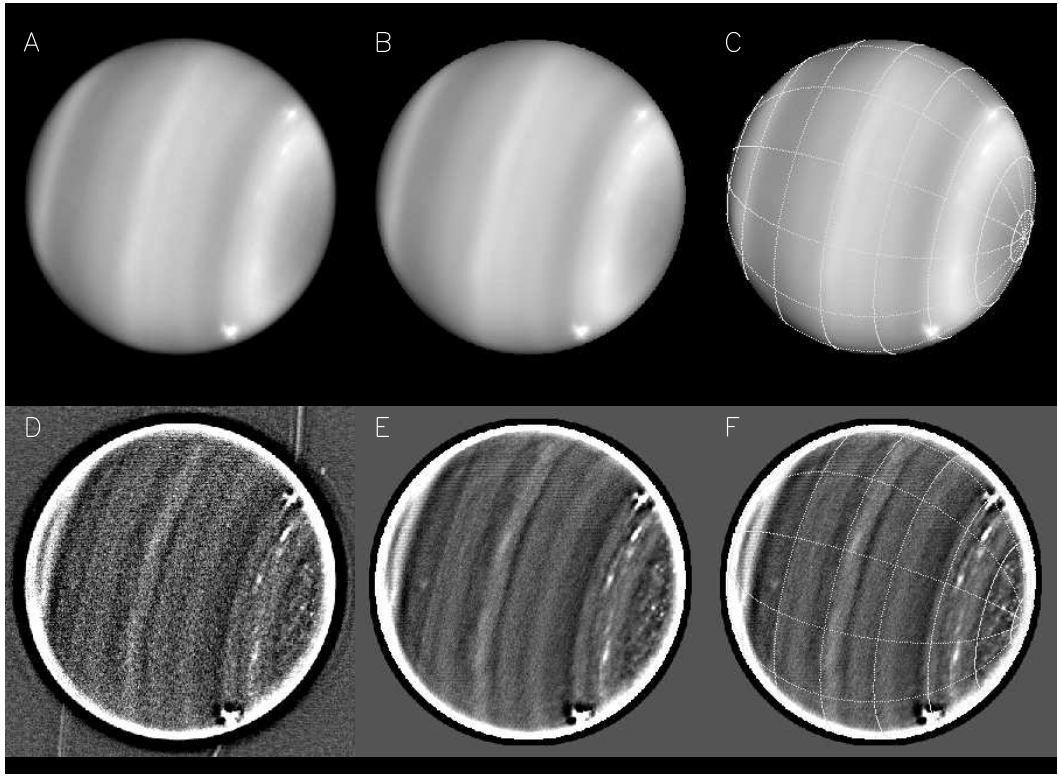}
\caption{Keck II NIRC2 H-filter image (2-minute exposure) from 26 July
  2011 (A) compared to an 8-image average with planet rotation removed
  (B). High-pass filtered versions (D) and (E), obtained by
  subtracting a 15-pixel $\times$ 15-pixel boxcar smoothed version from
  each, reveal a 2.6:1 improvement in S/N ratio.  The latitude grids in C and F
  are at 20\deg intervals starting at the equator. The longitude
grids are at 30\deg intervals.}
\label{Fig:kecksn}
\end{figure*}

Most of our post equinox observations are derived from data sets
designed to provide high S/N images.  The potential of this approach,
and results obtained from 2009 HST WFC3 imaging, are described by
\cite{Fry2012}.  Briefly, the approach is to take exposures short
enough to avoid significant smear due to planetary rotation during the
exposure, then average eight or more exposures together on a
latitude-longitude grid to remove the effects of planetary
rotation. Near the limb of the planet, images with better views are
given greater weight: we required the view angle cosine to be greater
than 0.025 and weighted each point by the square of the cosine. This
averaging process allows the detection of more subtle cloud features
than would otherwise be possible.  Sample results are illustrated in
Fig.\ \ref{Fig:kecksn}, where a single H image (A) is compared with an
8-image average (B, C).  Little difference is seen in the direct
images because the latitudinal variations in brightness are far
greater than the noise levels in both images.  But when the
large-scale latitudinal variations are removed by subtracting a
15-pixel $\times$ 15-pixel boxcar smoothed version of each image, the
improved S/N level of the averaged image becomes quite obvious, as
well as the improved visibility of subtle details in the cloud
structure. The measured central-disk noise level of the single image
is about 0.44\% of signal, and improves to about 0.17\% in the 8-image
average.  This improvement is about a factor of 2.6, which is
slightly below the factor of 2.83 ($\sqrt8$) expected for completely
random noise.

Except for the image S/N enhancement described above and in more
detail by \cite{Fry2012}, image processing and navigation followed the
same procedures described by \cite{Sro2009eqdyn}. We used standard
1-bar polar and equatorial radii of 24,973 km and 25,559 km
respectively, and a longitude system based on a 17.24-hour rotation
period \citep{Seidelmann2002}. 

\section{Cloud tracking}

\subsection{Methodology}

Initially, wind vectors were obtained via manual tracking of discrete
cloud features in high-pass filtered images, following the same
general approach described by \cite{Sro2009eqdyn}. Measurements of
longitude and latitude vs. time were fit to straight lines using both
unweighted regressions.  We used the weighted fit for the wind
estimate and the larger of weighted and unweighted error estimates for
the assigned error.  Errors in latitude and longitude were initially
computed assuming an angular error of 0.6 image pixels, based on
\cite{Sro2009eqdyn}. These estimated errors vary with view angle and
position on the disk, and are important when the number of target
measurements is small.  But when the number of samples is large, the
RMS deviation of the measurements from a straight line fit is a better
measure of the errors, as it includes both navigation errors as well
as target measurement errors. We find that tracking errors are
generally larger in longitude than in latitude because cloud
brightness gradients are generally much larger along meridians than
along lines of constant latitude. The main exceptions to this rule are
the small symmetrical spots seen in the north polar region of Uranus.

We also made extensive use of maximum correlation tracking to try to
increase the number of cloud targets and to reduce the errors in
defining their positions. To implement this approach we displayed an
image sequence as a stacked series of narrow horizontal strips, each
containing an orthogonal projection covering a specified range of
longitudes and a narrow range of latitudes.  For each cloud target
visible in the selected latitude range, a reference image is selected
and a target box is adjusted in size and position so that it contains
the cloud feature and a small region outside of it.  Target boxes in
other images are initially positioned using the 2007 wind
measurements, then manually adjusted in images in which the default
position fails to contain the target feature.  The positions of the
target boxes in all but the reference image are then automatically
refined to maximize the cross correlation between the reference target
box signal variations and those contained in each of the other
boxes. To reduce the impact of large-scale variations such as produced
by latitude bands, we use high-pass filters or median image
subtraction.  The correlation tracking usually takes a few iterations
to achieve convergence.  This procedure facilitated the identification
of more cloud targets, as well as providing generally more accurate
tracking results than purely manual measurements. For compact bright
features of relatively high contrast we also found it sometimes useful
for cloud tracking to use the center-of-differential brightness
coordinates $x_{CDB}$ and $y_{CDB}$, which are given by the
sums \begin{eqnarray} x_{CDB} = \sum_{i,j}^{target} x_i\cdot
  [I(i,j)-I_B(i,j)]/\\ \nonumber \sum_{i,j}^{target}
  [I(i,j)-I_B(i,j)]\\ y_{CDB} = \sum_{i,j}^{target}
  y_j\cdot[I(i,j)-I_B(i,j)]/\\ \nonumber \sum_{i,j}^{target}
  [I(i,j)-I_B(i,j)]\label{Eq:cdb}
\end{eqnarray}
where $I(i,j)$ is the reflectivity at image location $i,j$, $I_B(i,j)$
is either the smoothed reflectivity or the median filtered image
reflectivity, and the summation is over the area of the target box.
What we see in the filtered or subtracted images is $I-I_B$ rather
than $I$.

\subsection{July 2011 Keck results}

In our July 2011 Keck NIRC2 images we were able to identify 21 targets
suitable for tracking the winds of Uranus.  Most of these were only
trackable within a single transit on 26 July 2011, when weather and
seeing conditions did not severely impact the NIRC2 images.  On the
27th, turbulence increased significantly during the second half of the
night, when our observations were taken, and seeing reached extremely
poor levels up to several arc seconds FWHM, at which point the AO
system image quality deteriorates drastically.  Poor
seeing was also the rule on 28 July, although it was highly variable
and a few usable images were obtained.  But it was only the 26 July
images that could be combined together to produce enhanced
signal-to-noise ratios, which facilitated tracking of small
low-contrast cloud features.

\begin{figure}[!htb]\centering
\includegraphics[width=3.2in]{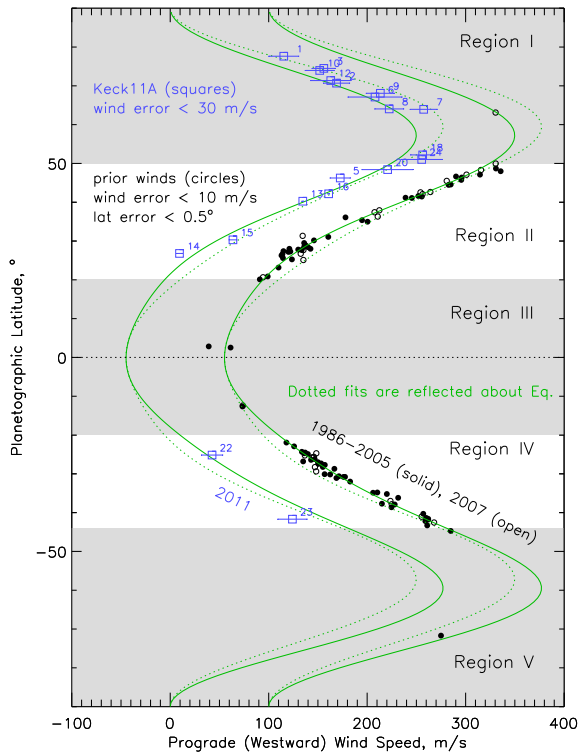}
\caption{Keck II wind measurements from July 26-28 2011 (squares) with
  errors less than 30 m/s, compared to the 10-term Legendre polynomial
  fit of \cite{Sro2009eqdyn} (solid curves) and its inverse about
  the equator (dotted curves).  Prior results are shown offset 100 m/s
  to the right.  Numeric labels refer to cloud target IDs given in
  Table\ \ref{Tbl:keckwinds}.  Shaded regions (I, III, and V) are
those poorly sampled by prior observations.}
\label{Fig:keckwinds}
\end{figure}

Even the 26 July images were far below the excellent quality
frequently obtained in prior observations, and thus it was not
surprising that the expected bounty of low contrast features implied
by the proof-of-concept image of \cite{Fry2012} did not
materialize. Not only was seeing below par (rapidly varying from
0.6$''$ to 1.2$''$ and beyond), but PSF characteristics (unusual wing structure
not normally visible) suggested that the AO system might have been out
of adjustment as well (this was later confirmed during the November
observing run of de Pater and Hammel).  Nevertheless, our imaging
program design partially compensated for these problems.  We were able
to make a significant improvement in the definition of the northern
hemisphere prograde jet, as illustrated in Fig.\ \ref{Fig:keckwinds}.
The wind accuracy we obtained for targets we could track was quite
impressive, often reaching uncertainties below 10 m/s. These new
manual tracking results are roughly consistent with the 10-term
Legendre polynomial fit of \cite{Sro2009eqdyn}, which is
based on wind measurements acquired during the intensive observing
period near the 2007 equinox.  We also obtained excellent views of
high northern latitudes that we present in a later section.

\begin{table*}\centering
\caption{Keck II NIRC2 wind results from 26-28 July 2011.}
\vspace{0.15in}
\begin{tabular}{c c r c c r r r r}
\hline
PC Lat.         & PG Lat.   &    Longitude & MJDavg   & Drift rate  & Zonal wind          &   ID   & $N$ & $\Delta t$ \\
(\degx ) & (\degx ) & (\deg east) & -55000  & (\degx/h east)            & (m/s west)          &        &           & (h)        \\
\hline
 77.07 &  77.64 $\pm$ 0.08 &  -4.7 $\pm$ 0.5 &  55768.5311 & -4.25 $\pm$ 0.56 &  115.2 $\pm$  15.4 &    1 &   13 &     3.02\\
 73.81 &  74.51 $\pm$ 0.11 &  81.9 $\pm$ 0.3 &  55768.5758 & -4.61 $\pm$ 0.35 &  155.8 $\pm$  12.0 &    3 &   10 &     2.65\\
 73.24 &  73.96 $\pm$ 0.17 & 313.6 $\pm$ 0.7 &  55770.5681 & -4.34 $\pm$ 0.42 &  151.9 $\pm$  14.9 &   10 &    3 &     3.98\\
 70.65 &  71.46 $\pm$ 0.12 & 120.5 $\pm$ 0.5 &  55768.6065 & -5.68 $\pm$ 0.76 &  228.6 $\pm$  31.2 &   11 &    5 &     1.83\\
 70.55 &  71.36 $\pm$ 0.12 & 106.4 $\pm$ 0.4 &  55768.5973 & -4.03 $\pm$ 0.51 &  163.0 $\pm$  20.9 &   12 &    6 &     2.24\\
 69.85 &  70.70 $\pm$ 0.10 &  -5.0 $\pm$ 0.3 &  55768.5263 & -4.04 $\pm$ 0.31 &  168.9 $\pm$  13.4 &    2 &    9 &     3.02\\
 67.14 &  68.08 $\pm$ 0.11 &  88.9 $\pm$ 0.3 &  55768.5883 & -4.52 $\pm$ 0.30 &  213.4 $\pm$  14.4 &    9 &    7 &     2.65\\
 66.14 &  67.11 $\pm$ 0.03 &  97.0 $\pm$ 0.3 &  55768.6186 & -4.23 $\pm$ 0.56 &  208.0 $\pm$  27.7 &    6 &    5 &     1.41\\
 62.99 &  64.05 $\pm$ 0.11 &  69.4 $\pm$ 0.2 &  55768.5975 & -4.03 $\pm$ 0.26 &  222.4 $\pm$  14.8 &    8 &    6 &     2.24\\
 62.87 &  63.93 $\pm$ 0.11 &  78.3 $\pm$ 0.2 &  55768.5746 & -4.64 $\pm$ 0.25 &  257.5 $\pm$  13.8 &    7 &    9 &     3.06\\
 50.91 &  52.20 $\pm$ 0.05 &  32.3 $\pm$ 0.1 &  55768.5526 & -3.33 $\pm$ 0.16 &  256.5 $\pm$  12.2 &   18 &   11 &     4.41\\
 49.71 &  51.02 $\pm$ 0.05 &  17.5 $\pm$ 0.2 &  55768.5398 & -3.23 $\pm$ 0.26 &  255.4 $\pm$  20.9 &   24 &    7 &     4.02\\
 47.09 &  48.41 $\pm$ 0.08 & 321.9 $\pm$ 0.2 &  55768.5077 & -2.65 $\pm$ 0.31 &  220.8 $\pm$  26.0 &   20 &    6 &     2.17\\
 44.92 &  46.25 $\pm$ 0.07 &  -6.8 $\pm$ 0.1 &  55768.5252 & -1.99 $\pm$ 0.12 &  172.7 $\pm$  10.2 &    5 &   16 &     3.44\\
 43.45 &  44.78 $\pm$ 0.04 & 124.5 $\pm$ 0.2 &  55768.6065 & -2.35 $\pm$ 0.38 &  209.1 $\pm$  33.4 &   19 &    5 &     1.83\\
 40.86 &  42.18 $\pm$ 0.08 & 269.2 $\pm$ 0.2 &  55769.2099 & -1.73 $\pm$ 0.01 &  160.9 $\pm$   0.9 &   16 &    7 &    48.19\\
 38.97 &  40.27 $\pm$ 0.07 &  72.0 $\pm$ 0.1 &  55768.7261 & -1.41 $\pm$ 0.01 &  134.5 $\pm$   0.9 &   13 &   12 &    52.17\\
 29.17 &  30.31 $\pm$ 0.15 & 198.7 $\pm$ 0.5 &  55770.1698 & -0.59 $\pm$ 0.04 &   63.7 $\pm$   3.7 &   15 &    3 &    21.18\\
 25.75 &  26.81 $\pm$ 0.07 & 128.8 $\pm$ 0.1 &  55768.7605 & -0.08 $\pm$ 0.03 &    9.4 $\pm$   2.9 &   14 &    7 &    24.01\\
-24.16 & -25.16 $\pm$ 0.09 &  19.4 $\pm$ 0.1 &  55768.5252 & -0.38 $\pm$ 0.09 &   42.5 $\pm$  10.5 &   22 &    8 &     3.01\\
-40.37 & -41.69 $\pm$ 0.15 &  29.0 $\pm$ 0.1 &  55768.5315 & -1.33 $\pm$ 0.16 &  124.1 $\pm$  14.8 &   23 &    6 &     2.50\\
\hline
\hline
\end{tabular}\vspace{0.07in}
\parbox{5.2in}{NOTES: PC Lat. and PG Lat. are planetocentric and planetographic latitudes respectively. MJDavg 
denotes Modified Julian Date average for the N measurements used to determine the cloud motion, where
MJD = Julian Date - 2\,400\,000.5. $\Delta$t is the time difference between the earliest and latest image
used to track a given target.}
\label{Tbl:keckwinds}
\end{table*}

\begin{figure*}[!htb]\centering
\includegraphics[width=5in]{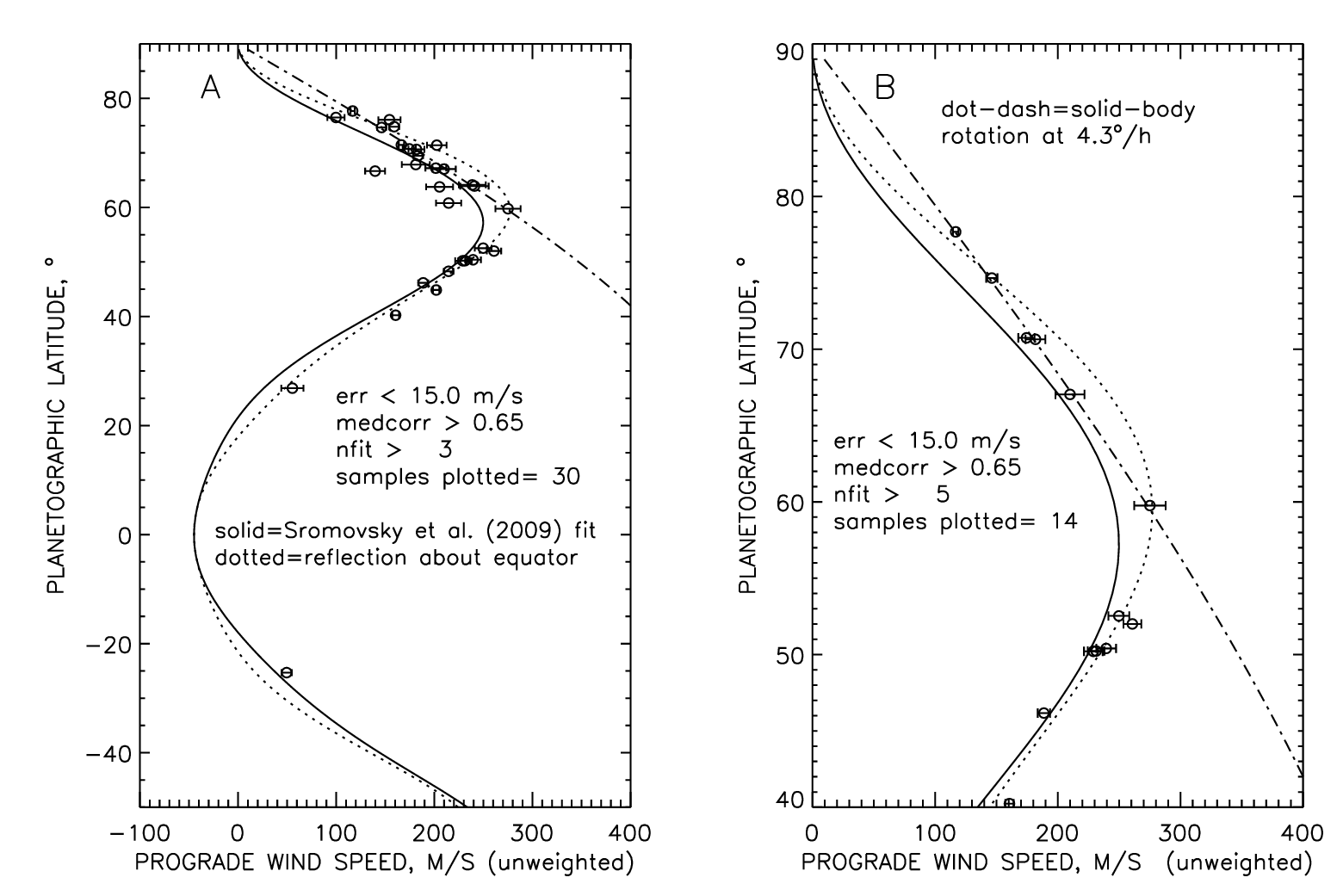}
\caption{Winds obtained from automated (mainly maximum-correlation)
  cloud tracking in 26 July 2011 Keck imagery for all latitudes and
  the best 30 vectors (A), and for high latitudes and the best 14
  vectors (B). The dot-dashed line in each figure indicates solid-body
rotation (latitudinally invariant rate of change of longitude with time)
at a rate of 4.3\degx/h westward.  In the legend err denotes wind
speed error, nfit is the number of points defining the motion of a
given cloud target, and medcorr is the median correlation coefficient
between the reference cloud target box and all the other target
boxes at their optimum displacements relative to the reference.}
\label{Fig:keckcorr}
\end{figure*}

We reanalyzed the Keck data set using the stacked remapped strips and
manually assisted cross-correlation method.  This enabled us to
identify 40 cloud targets in the 26 July data set alone.  Because only
the prominent cloud features can be reliably identified after more
than one planet rotation, we did not attempt to use these
correlation-based techniques on multiple days simultaneously.  For
such features our manual tracking already provides very high accuracy.
The results from the maximum-correlation tracking are displayed in
Fig.\ \ref{Fig:keckcorr} and Table\ \ref{Tbl:keckcorr}.  The most
accurate of the high latitude observations are shown in
Fig.\ \ref{Fig:keckcorr}B. These are filtered to have low formal error
estimates, high median correlation, and long tracking intervals. (Each
cloud target has a reference box containing the target in one image
that serves as a reference; the median correlation is the median of
the cross-correlation maximum values between the reference and the
subimage boxes in the other images.)  These results highlight small
but significant deviations from the \cite{Sro2009eqdyn} 10-term
Legendre polynomial fit. In the north polar region in particular the
polynomial fit has the wind speed and the longitudinal drift rate both
approach zero towards the pole, while our new measurements suggests
that the polar region is moving with solid body rotation at a drift
rate near 4.3\degx/h westward (the longitudinal drift rate is
independent of latitude for solid-body rotation).

\begin{table*}\centering
\caption{Correlation tracking results from 26 July 2011 Keck II images.}
\begin{tabular}{c c r c c r r r r}
\hline
PC Lat.         & PG Lat.   & Longitude & MJDavg   & Drift rate  & Zonal wind          &   ID   & $N$ & $\Delta t$ \\
(\degx ) & (\degx ) & (\deg east) & -55000  & (\degx/h east)            & (m/s west)          &     &        & (h)        \\
\hline
 77.67 &  78.21$\pm$0.25 & 122.3$\pm$0.3 & 768.6211 & -3.51$\pm$0.44 &   90.9$\pm$11.3 &  706 & 3 &  1.41\\
 77.12 &  77.69$\pm$0.15 &  -6.2$\pm$0.3 & 768.5352 & -4.32$\pm$0.13 &  116.7$\pm$3.4 &  701 & 8 &  4.04\\
 76.07 &  76.68$\pm$0.36 & -36.6$\pm$0.8 & 768.4883 & -5.48$\pm$0.92 &  159.9$\pm$26.8 &  707 & 4 &  1.37\\
 75.88 &  76.50$\pm$0.29 &   5.5$\pm$0.5 & 768.5039 & -3.37$\pm$0.30 &   99.7$\pm$8.7 &  702 & 5 &  2.60\\
 75.45 &  76.08$\pm$0.13 & -12.3$\pm$0.3 & 768.4883 & -5.12$\pm$0.47 &  155.9$\pm$14.3 &  703 & 4 &  1.37\\
 74.12 &  74.81$\pm$0.13 &  83.4$\pm$0.3 & 768.5938 & -4.79$\pm$0.15 &  159.1$\pm$5.1 &  705 & 5 &  2.65\\
 73.97 &  74.66$\pm$0.17 &  25.6$\pm$0.6 & 768.5469 & -4.37$\pm$0.14 &  146.2$\pm$4.6 &  704 & 9 &  4.43\\
 70.64 &  71.46$\pm$0.13 & 110.9$\pm$0.1 & 768.5938 & -4.20$\pm$0.13 &  168.8$\pm$5.1 &  714 & 5 &  2.65\\
 70.60 &  71.42$\pm$0.00 & 126.7$\pm$0.3 & 768.6055 & -5.02$\pm$0.25 &  202.5$\pm$9.9 &  715 & 4 &  1.83\\
 69.90 &  70.74$\pm$0.16 &  -7.0$\pm$0.4 & 768.5195 & -4.18$\pm$0.16 &  174.5$\pm$6.6 &  708 & 7 &  3.02\\
 69.80 &  70.65$\pm$0.24 &  43.2$\pm$0.6 & 768.5703 & -4.34$\pm$0.19 &  181.8$\pm$8.1 &  719 & 6 &  3.49\\
 68.60 &  69.49$\pm$0.07 &   4.4$\pm$0.3 & 768.5234 & -4.14$\pm$0.12 &  183.4$\pm$5.1 &  709 & 5 &  3.02\\
 66.92 &  67.86$\pm$0.22 &  91.5$\pm$0.6 & 768.5938 & -3.80$\pm$0.30 &  181.1$\pm$14.3 &  713 & 5 &  2.65\\
 66.37 &  67.33$\pm$0.25 &  77.8$\pm$0.6 & 768.5625 & -4.65$\pm$0.18 &  226.6$\pm$8.9 &  720 & 7 &  3.91\\
 66.24 &  67.20$\pm$0.18 & 103.7$\pm$0.5 & 768.5898 & -4.12$\pm$0.22 &  201.5$\pm$10.6 &  740 & 5 &  3.06\\
 66.07 &  67.04$\pm$0.20 &  16.2$\pm$0.4 & 768.5117 & -4.26$\pm$0.24 &  210.0$\pm$11.8 &  710 & 6 &  2.60\\
 65.68 &  66.66$\pm$0.10 & -44.7$\pm$0.2 & 768.4883 & -2.84$\pm$0.36 &  142.1$\pm$18.2 &  711 & 4 &  1.37\\
 64.35 &  65.37$\pm$0.17 &  44.6$\pm$0.4 & 768.6055 & -3.74$\pm$0.31 &  197.0$\pm$16.6 &  718 & 4 &  1.83\\
 63.08 &  64.14$\pm$0.13 &  73.7$\pm$0.3 & 768.6055 & -4.33$\pm$0.25 &  238.7$\pm$13.7 &  716 & 4 &  1.83\\
 62.85 &  63.91$\pm$0.13 &  80.1$\pm$0.3 & 768.6055 & -4.34$\pm$0.26 &  240.9$\pm$14.4 &  717 & 4 &  1.83\\
 62.70 &  63.77$\pm$0.18 & -42.2$\pm$0.2 & 768.4883 & -3.67$\pm$0.35 &  205.0$\pm$19.3 &  712 & 4 &  1.37\\
 59.83 &  60.97$\pm$0.27 &  10.7$\pm$0.5 & 768.5117 & -3.92$\pm$0.28 &  239.8$\pm$16.9 &  722 & 6 &  2.60\\
 59.66 &  60.81$\pm$0.25 & -10.4$\pm$0.2 & 768.5000 & -3.49$\pm$0.21 &  214.6$\pm$12.8 &  721 & 5 &  1.78\\
 58.59 &  59.76$\pm$0.20 &  63.4$\pm$0.7 & 768.5352 & -4.33$\pm$0.20 &  275.1$\pm$12.8 &  723 & 8 &  4.04\\
 51.63 &  52.92$\pm$0.06 & 126.4$\pm$0.3 & 768.6211 & -4.01$\pm$0.41 &  304.3$\pm$30.9 &  734 & 3 &  1.41\\
 51.30 &  52.59$\pm$0.20 & 135.9$\pm$0.2 & 768.6211 & -3.25$\pm$0.27 &  248.1$\pm$20.9 &  733 & 3 &  1.41\\
 51.24 &  52.53$\pm$0.31 &  25.4$\pm$0.5 & 768.5469 & -3.27$\pm$0.11 &  249.8$\pm$8.5 &  728 & 9 &  4.43\\
 50.71 &  52.01$\pm$0.21 &  30.2$\pm$0.3 & 768.5352 & -3.37$\pm$0.09 &  260.8$\pm$7.3 &  724 & 8 &  4.04\\
 49.09 &  50.40$\pm$0.54 &  56.3$\pm$0.3 & 768.5547 & -2.99$\pm$0.10 &  239.5$\pm$8.0 &  727 & 7 &  3.91\\
 48.92 &  50.23$\pm$0.19 &  45.6$\pm$0.3 & 768.5469 & -2.88$\pm$0.09 &  231.3$\pm$6.8 &  725 & 9 &  4.43\\
 48.90 &  50.21$\pm$0.24 & -29.2$\pm$0.2 & 768.5117 & -2.87$\pm$0.13 &  230.9$\pm$10.7 &  730 & 6 &  2.60\\
 48.67 &  49.98$\pm$0.28 &  63.5$\pm$0.4 & 768.5625 & -2.82$\pm$0.13 &  228.1$\pm$10.6 &  726 & 7 &  3.91\\
 46.96 &  48.28$\pm$0.15 & -40.9$\pm$0.1 & 768.5000 & -2.60$\pm$0.18 &  216.8$\pm$14.9 &  731 & 5 &  1.78\\
 44.84 &  46.17$\pm$0.26 &  -8.2$\pm$0.1 & 768.5195 & -2.21$\pm$0.07 &  191.7$\pm$6.3 &  729 & 7 &  3.02\\
 43.53 &  44.86$\pm$0.05 & 127.0$\pm$0.0 & 768.6055 & -2.27$\pm$0.13 &  201.3$\pm$11.4 &  732 & 4 &  1.83\\
 40.77 &  42.09$\pm$0.06 & -62.1$\pm$0.0 & 768.4805 & -1.99$\pm$0.21 &  185.0$\pm$19.9 &  736 & 3 &  0.94\\
 38.93 &  40.24$\pm$0.21 &  78.5$\pm$0.2 & 768.5469 & -1.68$\pm$0.04 &  160.6$\pm$3.4 &  735 & 9 &  4.43\\
 25.80 &  26.86$\pm$0.14 & 129.9$\pm$0.1 & 768.6055 & -0.46$\pm$0.12 &   51.3$\pm$13.0 &  737 & 4 &  1.83\\
-24.30 & -25.31$\pm$0.19 &  20.0$\pm$0.1 & 768.5117 & -0.44$\pm$0.05 &   49.0$\pm$5.9 &  738 & 6 &  2.60\\
-40.72 & -42.04$\pm$0.23 &  29.1$\pm$0.4 & 768.5312 & -1.41$\pm$0.20 &  131.5$\pm$18.5 &  739 & 6 &  2.50\\
\hline
\end{tabular}
\label{Tbl:keckcorr}
\parbox{4.75in}{NOTES: Column headings are as defined in Table\ \ref{Tbl:keckwinds}.}
\end{table*}

\subsection{Gemini October 2011 results}

\begin{figure*}[!htb]\centering
\includegraphics[width=3in]{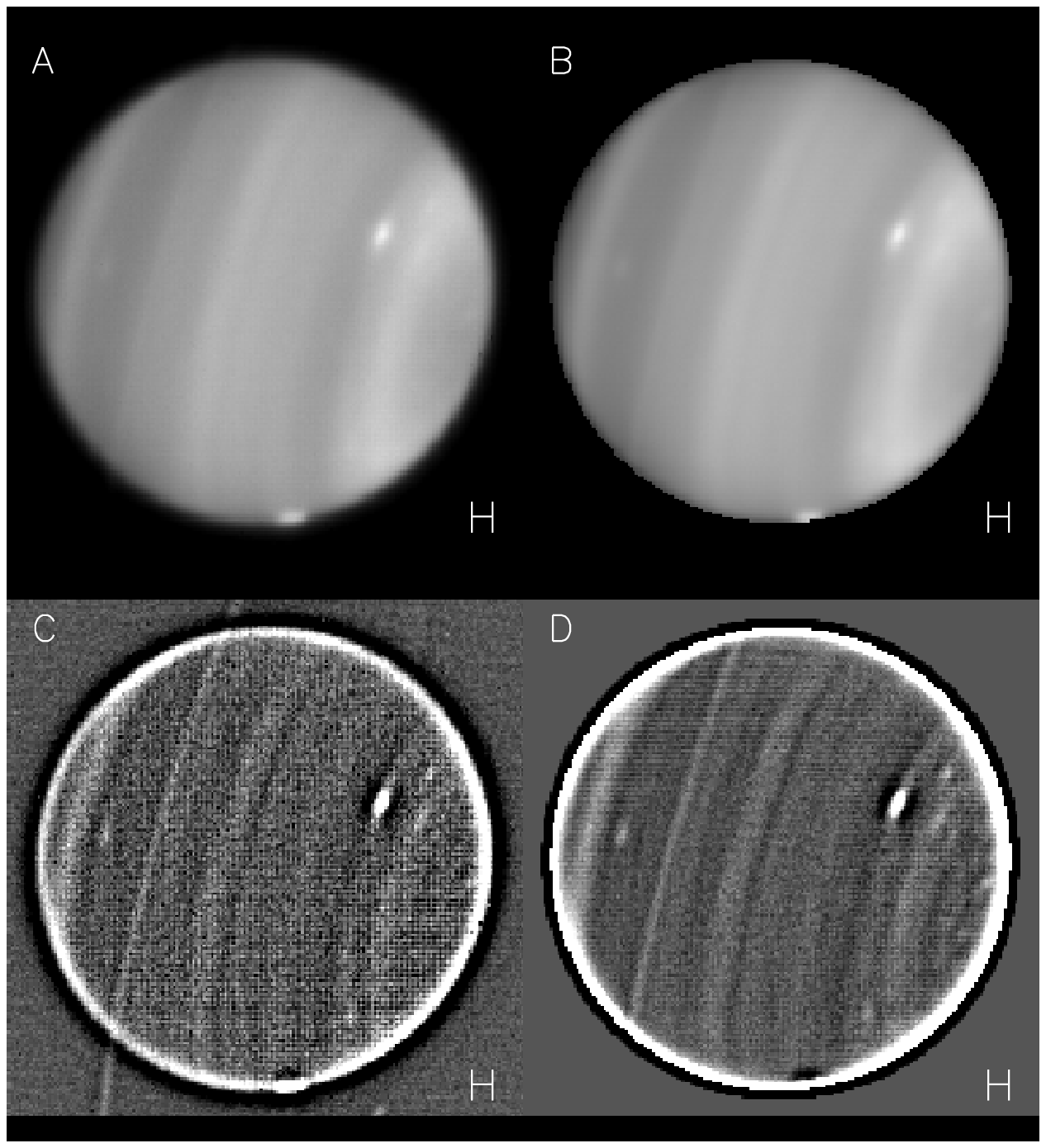}
\includegraphics[width=3in]{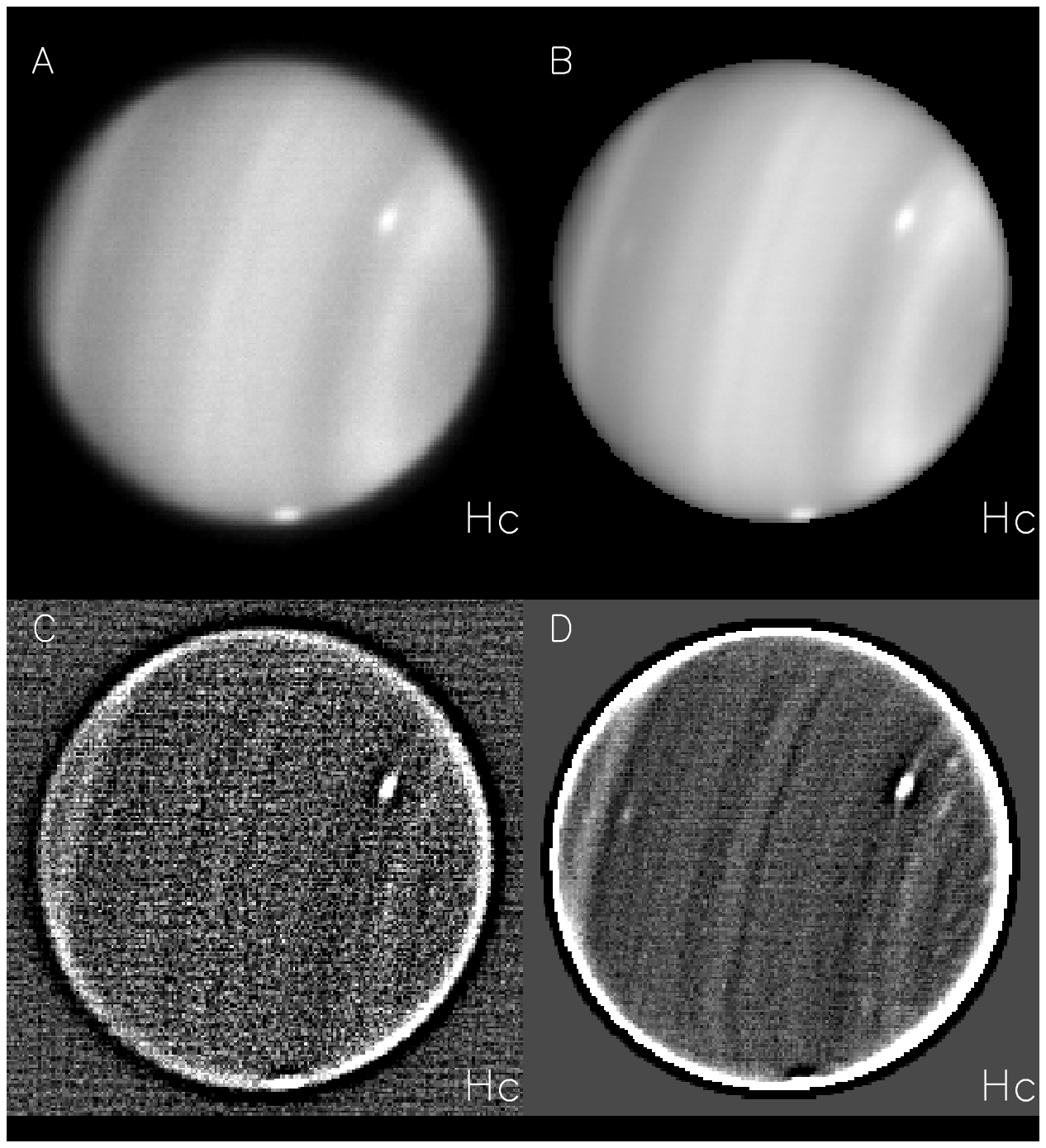}
\caption{Left: Gemini-North H filter image from 26 October 2011 (A) compared
  to an 8-image average with planet rotation removed (B). All
  individual images were taken with a 2-minute exposure. High-pass
  filtered versions (C) and (D), obtained by subtracting a 7-pixel
  $\times$ 7-pixel boxcar smoothed version from each, reveal a large
  improvement in S/N ratio. Right: Same as at left, except for the
  Hcont filter. }
\label{Fig:gemsn}
\end{figure*}

Gemini-North observations were obtained during 23, 25, and 26 October
2011 (Table\ \ref{Tbl:tracklist}), with most of the high-quality
imaging obtained during 26 October.  As with the Keck II imaging
in July, we used a program of repeated short exposures and averaging
on latitude-longitude grids to improve S/N dramatically and eliminate
smear due to planet rotation.  Sample improvements can be seen in
Fig.\ \ref{Fig:gemsn}, which compares single H and Hcont images
with 8-image average versions.

\begin{figure}[!htb]\centering
\includegraphics[width=3.3in]{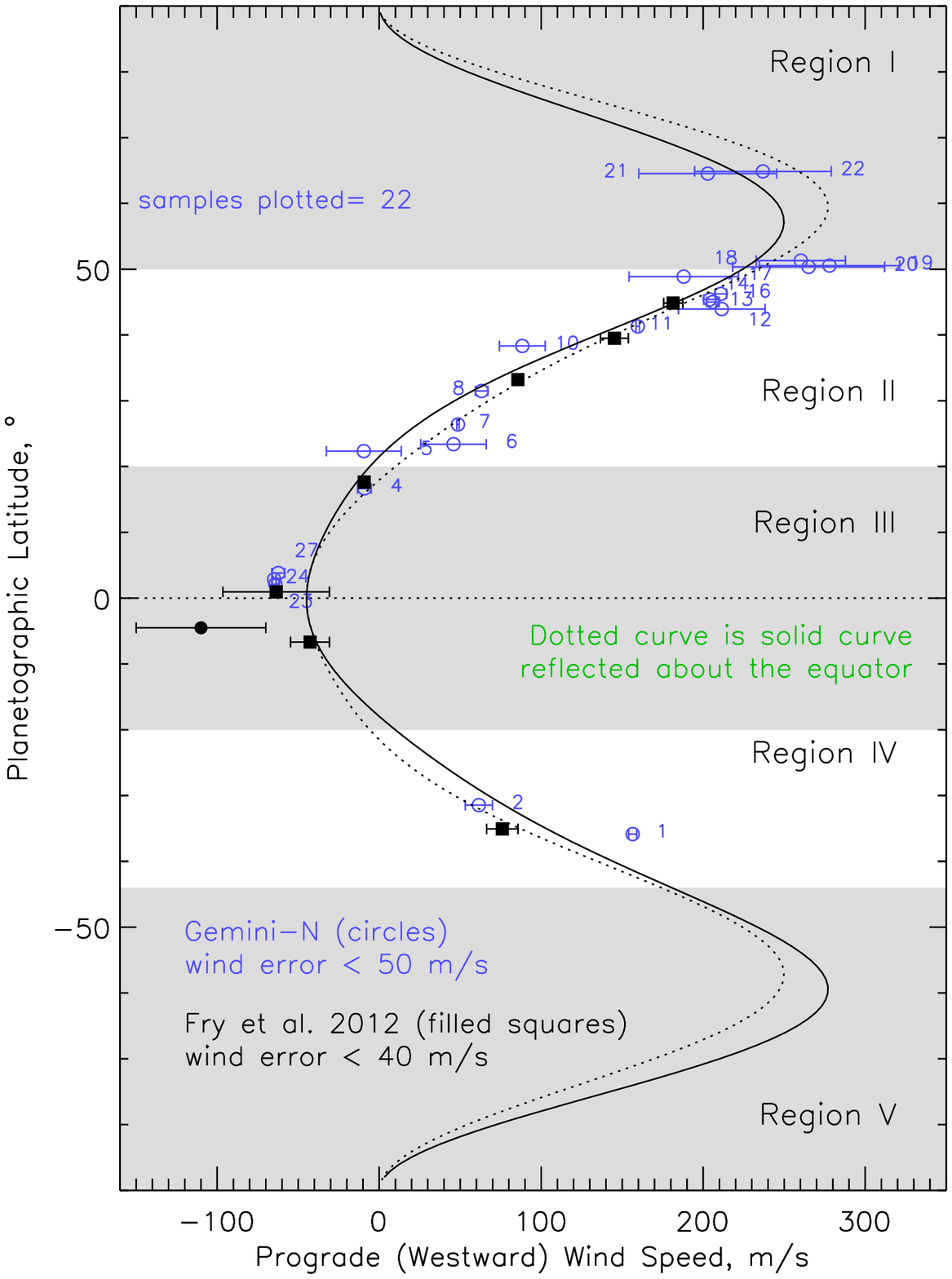}
\caption{Gemini-North wind measurements from October 23-26 2011 (circles) with
  errors less than 50 m/s, compared to the 10-term Legendre polynomial
  fit of \cite{Sro2009eqdyn} (solid curve) and its inverse about
  the equator (dotted curve).  Also plotted are 2009 HST results
from \cite{Fry2012} (filled squares) and the \cite{Lindal1987} radio
occultation result (filled circle).  Numeric labels refer to cloud target
  IDs given in Table\ \ref{Tbl:gemcorr}.}
\label{Fig:gemwinds}
\end{figure}

Using a time-sequence display in combination with manual selection and
tracking, we were able to track 21 cloud features; but only two were
tracked long enough to determine highly accurate wind speeds.  To
improve our analysis of this data set we applied our stacked strip and
cross-correlation tracking method.  This led to an increase in the
number of trackable features from 21 to 26, with 9 features tracked
over more than a single transit, compared to just two in the initial
manual effort.  These long time spans yield extremely accurate drift
rates and wind speeds, with uncertainties of less than 0.04\degx/h and
4 m/s respectively.  The results are listed in
Table\ \ref{Tbl:gemcorr} and a subset are plotted in
Fig.\ \ref{Fig:gemwinds}, where the wind results are compared to the
10-term Legendre fit of \cite{Sro2009eqdyn} and with the results of
\cite{Fry2012}, which are shown as filled squares.  The latter results
are in best agreement with the inverted Legendre fit, as is the most
accurate of our Gemini wind measurements. Although this weakly suggests that
the circulation may be in the process of seasonal change in the
direction of reversing the north-south asymmetry observed during the
last (2007) equinox, the number of samples supporting this
change is too small to consider statistically significant.  Most of
the Gemini results are at latitudes where we already have significant
numbers of samples, but Gemini and the \cite{Fry2012} results both
contribute vectors in Region III, which has been grossly undersampled
in most observations, due to a lack of cloud features visible at the
available S/N.

\begin{table*}\centering
\caption{Gemini-North correlation cloud tracking results from 23-26 October 2011.}
\begin{tabular}{r r r r r r r r r}
\hline
PC Lat.         & PG Lat.   &    Longitude & MJDavg   & Drift rate  & Zonal wind          &   ID   & $N$& $\Delta t$\\
(\degx ) & (\degx ) & (\deg east) & -55000  & (\degx /h  east)            & (m/s west)          &        &           & (h)        \\
\hline
63.83 &  64.87 $\pm$ 0.85 & 284.8 $\pm$ 1.1 &    860.2656 & -4.42 $\pm$ 0.79 &  236.9 $\pm$  42.2 &  222 &    6 &     2.02\\
 63.49 &  64.54 $\pm$ 0.71 & 308.0 $\pm$ 1.9 &    860.2852 & -3.74 $\pm$ 0.78 &  202.8 $\pm$  42.6 &  221 &    7 &     3.03\\
 50.00 &  51.30 $\pm$ 0.22 & 339.9 $\pm$ 0.9 &    860.3242 & -3.31 $\pm$ 0.35 &  260.3 $\pm$  27.5 &  218 &    9 &     3.03\\
 49.24 &  50.55 $\pm$ 0.39 & 244.5 $\pm$ 1.2 &    860.2578 & -3.48 $\pm$ 0.54 &  277.9 $\pm$  43.2 &  219 &   10 &     2.27\\
 49.04 &  50.35 $\pm$ 0.40 & 245.3 $\pm$ 1.3 &    860.2578 & -3.31 $\pm$ 0.58 &  265.1 $\pm$  46.9 &  220 &   10 &     2.27\\
 47.56 &  48.88 $\pm$ 0.36 & 202.9 $\pm$ 0.8 &    859.3359 & -2.28 $\pm$ 0.41 &  188.0 $\pm$  33.7 &  217 &    5 &     2.71\\
 44.92 &  46.25 $\pm$ 0.23 & 206.8 $\pm$ 0.8 &    859.6953 & -2.43 $\pm$ 0.04 &  210.9 $\pm$   3.5 &  216 &    5 &    21.30\\
 44.87 &  46.20 $\pm$ 0.30 & 247.2 $\pm$ 0.7 &    860.2383 & -1.38 $\pm$ 0.76 &  119.9 $\pm$  65.7 &  215 &    6 &     1.26\\
 44.06 &  45.39 $\pm$ 0.44 &  39.9 $\pm$ 0.9 &    860.1875 & -2.32 $\pm$ 0.04 &  204.0 $\pm$   3.6 &  214 &    7 &    26.60\\
 43.69 &  45.02 $\pm$ 0.41 &  39.7 $\pm$ 1.0 &    860.1875 & -2.33 $\pm$ 0.04 &  206.1 $\pm$   3.9 &  213 &    7 &    26.60\\
 42.62 &  43.95 $\pm$ 0.25 & -24.5 $\pm$ 1.0 &    860.3047 & -2.34 $\pm$ 0.30 &  211.5 $\pm$  26.7 &  212 &   13 &     3.03\\
 40.00 &  41.31 $\pm$ 0.35 & 152.2 $\pm$ 0.8 &    858.6875 & -1.70 $\pm$ 0.01 &  159.6 $\pm$   1.1 &  211 &   10 &    50.38\\
 37.05 &  38.34 $\pm$ 0.12 & 252.8 $\pm$ 0.4 &    860.2695 & -0.90 $\pm$ 0.14 &   88.3 $\pm$  14.1 &  210 &   12 &     2.77\\
 34.88 &  36.14 $\pm$ 0.36 & -15.9 $\pm$ 0.5 &    860.3477 & -0.24 $\pm$ 0.68 &   24.0 $\pm$  68.3 &  209 &    5 &     1.01\\
 30.32 &  31.49 $\pm$ 0.54 &  52.2 $\pm$ 0.7 &    860.1367 & -0.60 $\pm$ 0.04 &   63.3 $\pm$   3.9 &  208 &    5 &    26.60\\
 25.36 &  26.40 $\pm$ 0.17 & 131.8 $\pm$ 0.3 &    858.9336 & -0.44 $\pm$ 0.01 &   48.6 $\pm$   0.7 &  207 &   11 &    50.87\\
 22.45 &  23.40 $\pm$ 0.30 &   8.3 $\pm$ 0.2 &    860.3320 & -0.40 $\pm$ 0.18 &   45.9 $\pm$  20.2 &  206 &    8 &     1.77\\
 21.43 &  22.35 $\pm$ 0.31 &   7.7 $\pm$ 0.2 &    860.3398 &  0.08 $\pm$ 0.20 &   -9.5 $\pm$  23.1 &  205 &    7 &     1.51\\
 15.96 &  16.68 $\pm$ 0.33 & 183.5 $\pm$ 0.7 &    859.6953 &  0.08 $\pm$ 0.04 &   -9.1 $\pm$   4.2 &  204 &    5 &    21.30\\
  3.65 &   3.82 $\pm$ 0.10 & 201.4 $\pm$ 0.6 &    859.7773 &  0.50 $\pm$ 0.03 &  -62.3 $\pm$   4.0 &  227 &    4 &    21.30\\
  3.24 &   3.39 $\pm$ 0.13 & 331.1 $\pm$ 1.9 &    860.3203 &  0.32 $\pm$ 0.84 &  -39.4 $\pm$ 103.5 &  226 &    9 &     2.52\\
  2.74 &   2.87 $\pm$ 0.22 & 264.0 $\pm$ 0.8 &    859.3945 &  0.52 $\pm$ 0.01 &  -64.9 $\pm$   1.2 &  224 &    7 &    72.70\\
  2.27 &   2.38 $\pm$ 0.45 & 265.0 $\pm$ 1.2 &    860.2422 &  0.66 $\pm$ 0.93 &  -81.5 $\pm$ 114.7 &  225 &    6 &     1.76\\
  2.03 &   2.13 $\pm$ 0.34 & 242.1 $\pm$ 1.5 &    858.9688 &  0.52 $\pm$ 0.02 &  -64.1 $\pm$   2.2 &  223 &    7 &    72.95\\
-30.28 & -31.45 $\pm$ 0.25 & 285.4 $\pm$ 0.3 &    860.2852 & -0.58 $\pm$ 0.08 &   61.5 $\pm$   8.4 &  202 &   13 &     3.03\\
-34.60 & -35.85 $\pm$ 1.40 & 236.3 $\pm$ 0.3 &    860.0117 & -1.55 $\pm$ 0.02 &  156.5 $\pm$   2.5 &  201 &    4 &    20.61\\
\hline
\end{tabular}
\label{Tbl:gemcorr}
\parbox{4.85in}{NOTES: Column headings are as defined in Table\ \ref{Tbl:keckwinds}.}
\end{table*}

\subsection{Equatorial winds and cloud patterns}

\begin{figure*}\centering
\includegraphics[width=5in]{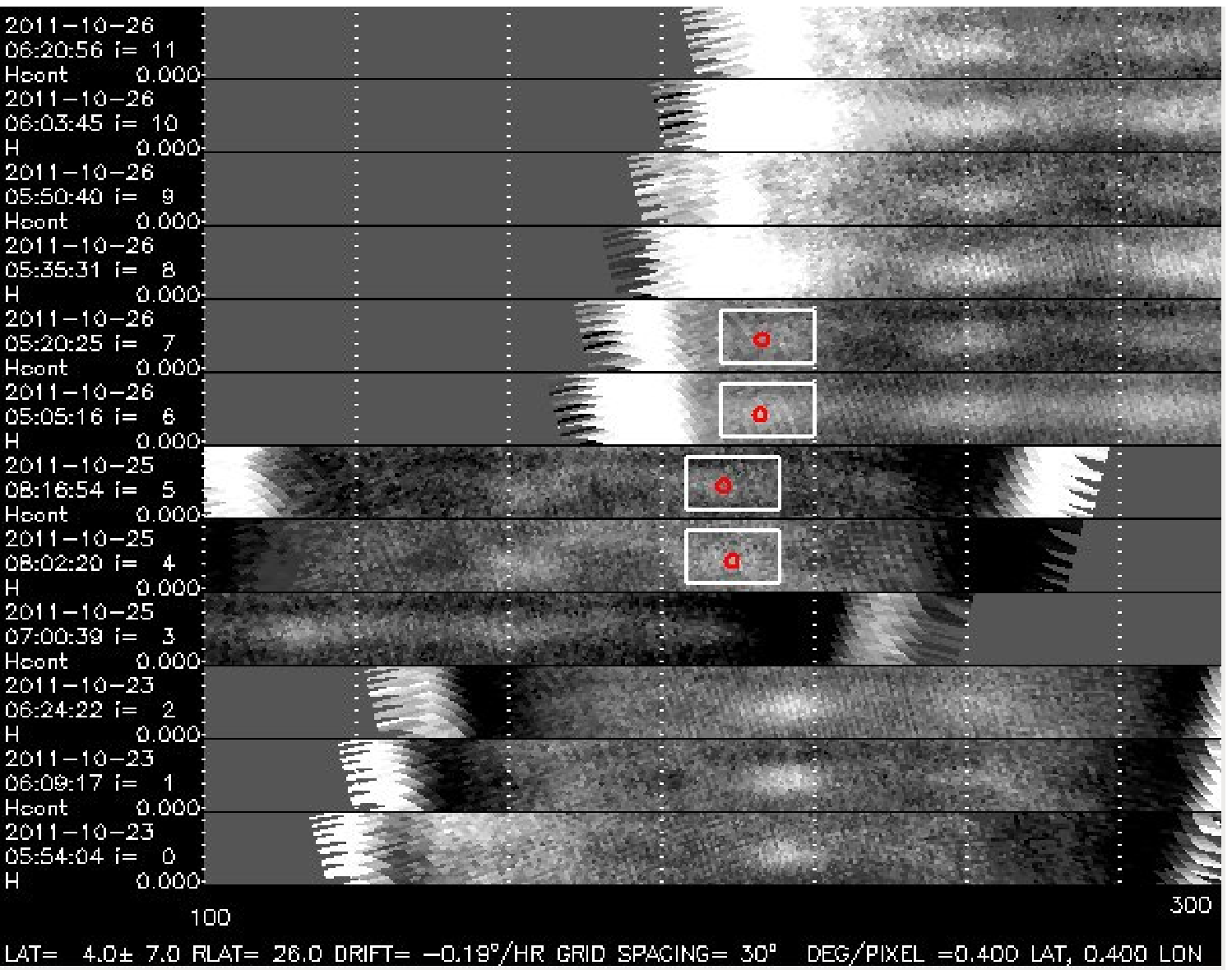}
\caption{Diffuse equatorial features illustrated in this
stack of rectilinear projections of median subtracted 2011 Gemini NIRI
images from 23 October (bottom three strips), 25 October (next three
strips), and 26 October (top six strips). The boxes outline
the same feature that appeared on successive rotations.  Also note the 
pattern of roughly 40\deg separation between equatorial features.}
\label{Fig:eqstack}
\end{figure*}

There are broad and fuzzy cloud patterns centered at approximately
4\deg N, which have a latitudinal width of 5-6\deg a longitudinal
extent of $\sim$8\degx, and a spacing of approximately 40\degx, which
corresponds to a wavenumber-9 pattern of broad cloud features.  A
similar but more complete pattern was captured by 2003 Keck imagery,
although in that case a wavenumber of 12 was inferred by
\cite{Hammel2005winds}. Given some variability in locating these
broad features, it is not clear whether these two patterns are really
different.  The 2011 equatorial features are most easily
seen in images that have had a median image subtracted.  A stack of
remapped difference images of that type are shown in
Fig.\ \ref{Fig:eqstack}.  Median subtraction has the effect of
removing the zonally symmetric and latitude dependent background I/F
variation, leaving only the additional brightness produced by the
discrete cloud features.  This is most effective when the images in
the data set all have the same resolution and image quality, and there
are sufficient numbers on different rotations that discrete features
do not contribute to the median image.

\begin{figure}\centering
\includegraphics[width=3.2in]{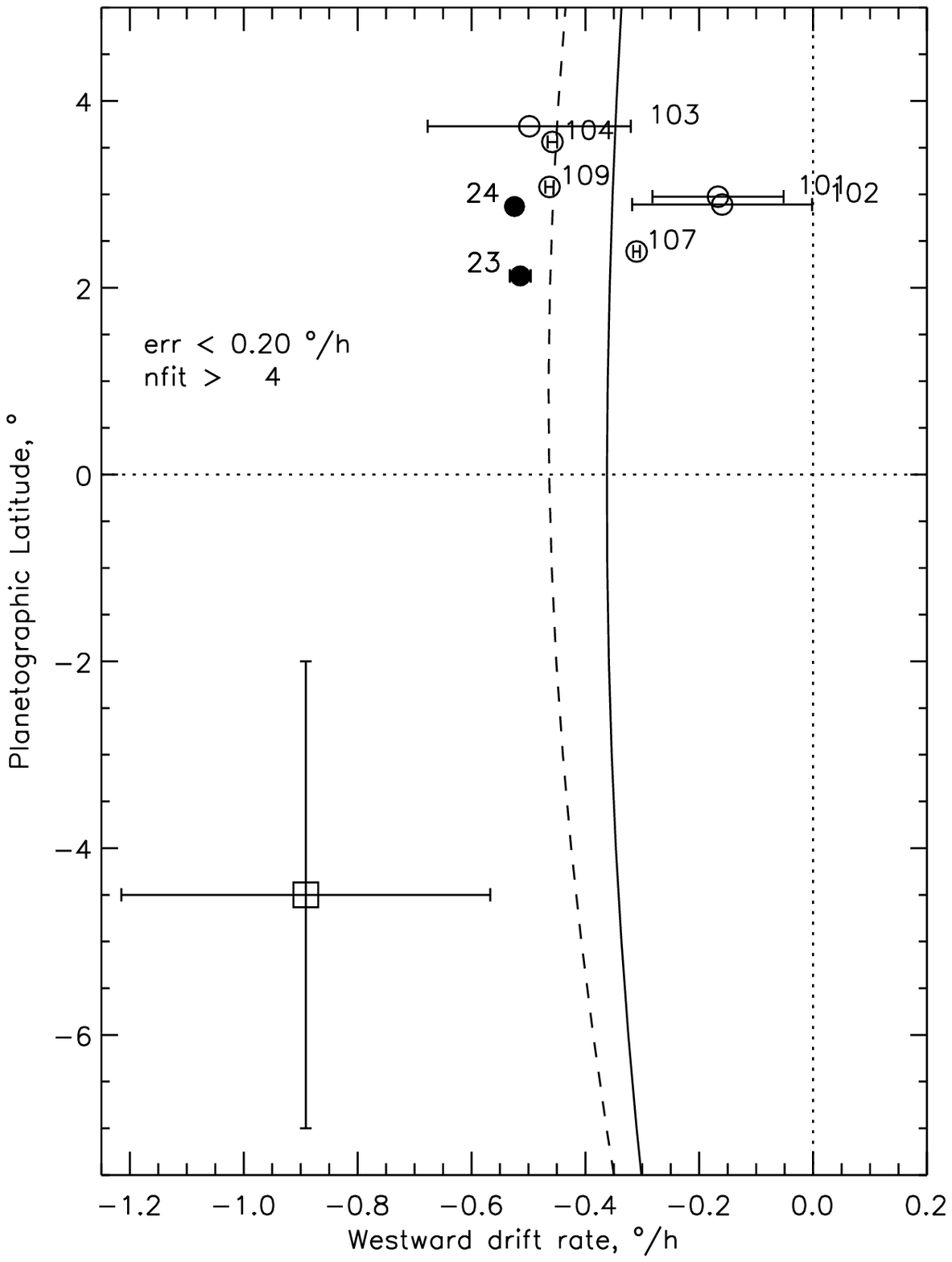}
\caption{Comparison of low-latitude wind measurements in 2011 Gemini images
(filled circles) with measurements in 2003 Keck images (open circles) and
the \cite{Lindal1987} radio occultation measurement (square), for which the
vertical error bar here represents the range of latitudes sampled by
ingress and egress measurements on which the wind determination is based.
The solid curve indicates the 10-term Legendre polynomial fit of \cite{Sro2009eqdyn},
while the dashed curve is our 13-term fit described in Section\ \ref{Sec:fits}.}
\label{Fig:eqwind}
\end{figure}

Tracking the motions of the equatorial features has been very
difficult because of their poorly defined boundaries, so that their
motions within a single transit cannot be well enough defined to
project their positions on a subsequent planet rotation with enough
accuracy to convincingly show which feature among several choices is
the same one seen on the previous rotation.  Fortunately, we found a
few cases (as illustrated in Fig.\ \ref{Fig:eqstack}) in which
neighboring rotations also contain a feature, which resolved the
ambiguity.  This has also been confirmed by tracking much sharper
features in 2003 Keck images of \cite{Hammel2005winds}, which were
taken on four successive nights and have relatively high and stable
image quality.  These two data sets make it quite clear that the
equatorial drift rate is relatively small ($\approx$0.5\degx/h
eastward) and has changed little between 2003 and 2011, as illustrated
in Fig.\ \ref{Fig:eqwind}.  \cite{Hammel2005winds} obtained similar
equatorial wind speeds, and pointed out that these speeds may
represent the speed of a wave feature rather than that of the mass
motion.  The only measurement of the mass flow is the Voyager 2 radio
occultation result of \cite{Lindal1987}, which is 110$\pm$40 m/s
eastward (0.89$\pm$0.32\degx/h). Although this is a rather uncertain
value, it does suggest that the equatorial features may be moving
slightly westward relative to the zonal mass flow.  On earth this
phase speed would suggest Rossby waves, but because Uranus is a
retrograde rotator, they would move eastward relative to the zonal
flow. The more plausible alternative is a Kelvin wave, although it is
not clear if the suggested wavenumber and phase speed magnitude are
compatible.  The magnitude itself is highly uncertain because of the
large uncertainty in the radio-derived wind speed.

\subsection{Long-lived feature tracking results}

A few cloud features on Uranus have extremely long lifetimes, and some
have been tracked for years.  These include a bright high-altitude
cloud feature seen near 30\deg N \citep{Sro2007bright} and the large
Berg feature.  The latter feature oscillated between 32\deg S and
36\deg S for many years, then began drifting northward in 2005
\citep{Sro2005dyn, Sro2009eqdyn, DePater2011} and dissipated as it
approached the equator.  Additionally, \cite{Sro2009eqdyn} were able
to track eight features for well over a month (1055-2250 hours) during
the intensive equinox observing period.  Earlier, \cite{Kark1998Sci}
had noted that over the 100 day period covered by his 1997 and 1998
NICMOS observations, all eight of the features he detected were
visible whenever they were on the sunlit side of Uranus.  Thus we
expect at least some of the larger features on Uranus to be trackable
over an extended time period.

 More recently, another long-lived bright feature near 30\deg N was
 tracked from 26 July 2011 until 16 December 2011
 \citep{Sro2012bs}.  This is the same feature we identified here with
 target ID 205 and 206 in the Gemini data set, and manual target ID 14
 in the July Keck data set (correlation target 737).  When the Keck
 and Gemini observations are combined together, the time baseline
 increases enormously and the mean drift rate can be determined with
 high accuracy.  The longitude changed from 128.8\degx E at JD
 55768.7605 to 8.3\degx E (the 206 value) at 55860.3320, a time
 difference of 91.5715 days (2197.72 h).  In the Keck data set the
 drift rate is estimated to be -0.08$\pm$0.03\degx/h, which implies a
 decrease in east longitude of 176$\pm$66\degx over that time
 interval.  Thus the predicted position in our October Gemini image is
 at -47$\pm$66\degx E.  The actual position is 8.3\degx E (mod
 360\degx).  Regarding the possibility of 360\deg multiples being
 added or subtracted, we find that the only plausible value within the
 prediction limits is 8.3\deg E.  This implies a drift rate of
 (8.3\degx E - 128.8\degx E)/2197.72 h = -0.0548\degx E/h
 (-1.316\degx /day), with an uncertainty of about 0.0007\degx/h
 (0.016\degx/day).  This highly accurate mean drift rate does not mean
 that the drift rate did not vary during the time between
 observations. In fact, \cite{Sro2012bs} show that the rate did vary
 after this period.

A second long-lived bright spot was also identified by
\cite{Sro2012bs} at a similar latitude.  This feature is present in
the July Keck data set, but not seen over a long enough time period
to be tracked within that data set. It was tracked in the Gemini data
set, and given ID 207, which then was at 26.40\deg N.  Our drift rate
of -0.44$\pm$0.01\degx E/h (or -10.56$\pm$0.24\degx E/day) is somewhat
larger than the long term averages of \cite{Sro2012bs} for this
feature, which ranged from -9.125 to -9.352 \degx E/day depending on the
period covered.  However, it does appear roughly consistent with the average
motion of the feature between October and November.

\subsection{HST and Gemini observations in 2010}

HST observations of Uranus were made in June, July, and August
2010. The June observations provide complete longitude coverage but
temporal sampling and the small number of discrete cloud features
prevents any wind determination from this data set alone.  The July
and August observations are part of a SNAP program, and provide
neither complete longitude coverage nor suitable temporal sampling for
internal wind determinations.  We also have November 2010 imaging
observations of Uranus from Gemini-North using the NIRI instrument.
These also have limited coverage, suffer from poor seeing generally,
and don't have adequate temporal sampling for internal wind
measurements.  However, relying on the long lifetime of most clouds on
Uranus, we can combine these data sets to obtain a small number of
very precisely determined drift rates. These data sets contain a few
cloud features in the northern hemisphere that we attempted to track,
even though time gaps were long.  However, none of the features could
be convincingly demonstrated to have survived over these long
intervals and to have had a uniform drift rate.  It was always
possible to find a drift rate solution when the number of observations
was small, but when the number of observations was large, a steady
drift solution could not be found.

\section{Legendre polynomial fits and symmetry properties}

\subsection{Fitting methodology}

To provide a smooth profile for use by atmospheric modelers and other
researchers, we carried out Legendre polynomial fits to combinations
of observations from 2009 and 2011, which includes 30 points, and to a
more extensive set of observations that also included the highest
accuracy observations from 2007, which contains an additional 28
points. We fit the angular (longitudinal) drift rates, rather than
wind speeds because the observations are most consistent with a
constant angular drift rate at high latitudes.  The longitudinal drift
rate does not change as dramatically at high latitudes as does the
wind speed, and thus is much easier to fit without generating violent
deviations at other latitudes where data are more sparse.  To limit
large deviations in the sparsely sampled regions, we also limited the
order of the polynomials to 11th order (12 terms) for the 2009-2011
data set, and to 13th order (14 terms) for the 2007-2011 data set. We
also added the 1986 Voyager measurement at 71\deg S
\citep{SmithBA1986} and artificial measurements of 4.3\degx/h westward
at both poles to better constrain high latitudes. The angular rate 4.3\degx/h
was chosen to be consistent with the nearest observed values shown in
Fig.\ \ref{Fig:legfit9-11}A.
The Voyager point
was not needed for fits that were constrained to be symmetric.
Because both of these fits are more symmetric than the fit we
obtained to the 2007 data alone, we also considered it possible that
the circulation of Uranus is actually close to symmetric, and carried
out fits in which only the even Legendre polynomials were used,
insuring fits that were symmetric about the equator.

We modeled longitudinal drift rates and
wind speeds  using the following Legendre expansion and conversion
equations:\begin{eqnarray} d\phi/dt = \sum_{i=0}^{n} C_i \times
  P_i(\sin(\theta))\\ U = 4.8481\times10^{-3} R(\theta) \times
  d\phi/dt \\ R(\theta) = R_E/\sqrt{1+(R_P/R_E)^2 \tan(\theta)^2}
\end{eqnarray}
where $C_i$ are the coefficients given in Table\ \ref{Tbl:legfits},
$P_i(\sin(\theta))$ is the $ith$ Legendre polynomial evaluated at the
sine of planetographic latitude $\theta$, $d\phi/dt$ is the westward
longitudinal drift rate in \degx /h, $U$ is the wind speed in m/s,
$R$ is the radius of rotation in km at latitude $\theta$, which is the
distance from a point on the 1-bar surface to the planet's rotational
axis, $R_E$ and $R_P$ are the equatorial and polar radii of Uranus.
For the symmetric fits, the summation over $i$ is only over the even
polynomials.  The model coefficients are found by minimizing $\chi^2$,
but with error estimates for the observations modified as described in
the following paragraph.

Because very accurate measurements of drift rates at nearly the same
latitude often did not agree within their uncertainties, and often
differed by many times the value expected from those uncertainties, it
is clear that one of the following possibilities must be considered:
(1) the circulation is not entirely steady, or (2) that the features
that we measure do not all represent the same atmospheric level, or
(3) the cloud features we track are not always at the same latitude as
the circulation feature that is moving with the zonal flow. Examples
of the latter possibility are the companion clouds to Neptune's Great
Dark Spot, which traveled with the spot even though separated by
nearly 10\deg in latitude \citep{Sro1993Icar}. Less extreme examples
have also been seen on Uranus \citep{Hammel2009Icar, DePater2011}.  If
we include highly accurate measurements, and weight them by their
estimated accuracy, they can dominate the fit, leading to unreasonable
variations in regions where there are less accurate measurements.
Since these high accuracy measurements clearly do not all follow the
mean flow, we must add an additional uncertainty to characterize their
deviations from the mean flow.  We do this by root sum squaring the
estimated error of measurement with an additional error of
representation.  We adjust the size of this error until the $\chi^2$
value of the complete asymmetric fit is approximately equal to the
number of degrees of freedom (number of measurements minus the number
of fitted parameters). This representative error (referred to as
reperr in the figure legends) turns out to be
approximately 0.1\degx/h for both the 2009-2011 and 2007-2011 data
sets.

\subsection{Fit results}\label{Sec:fits}

\begin{figure*}[!htb]\centering
\includegraphics[width=6in]{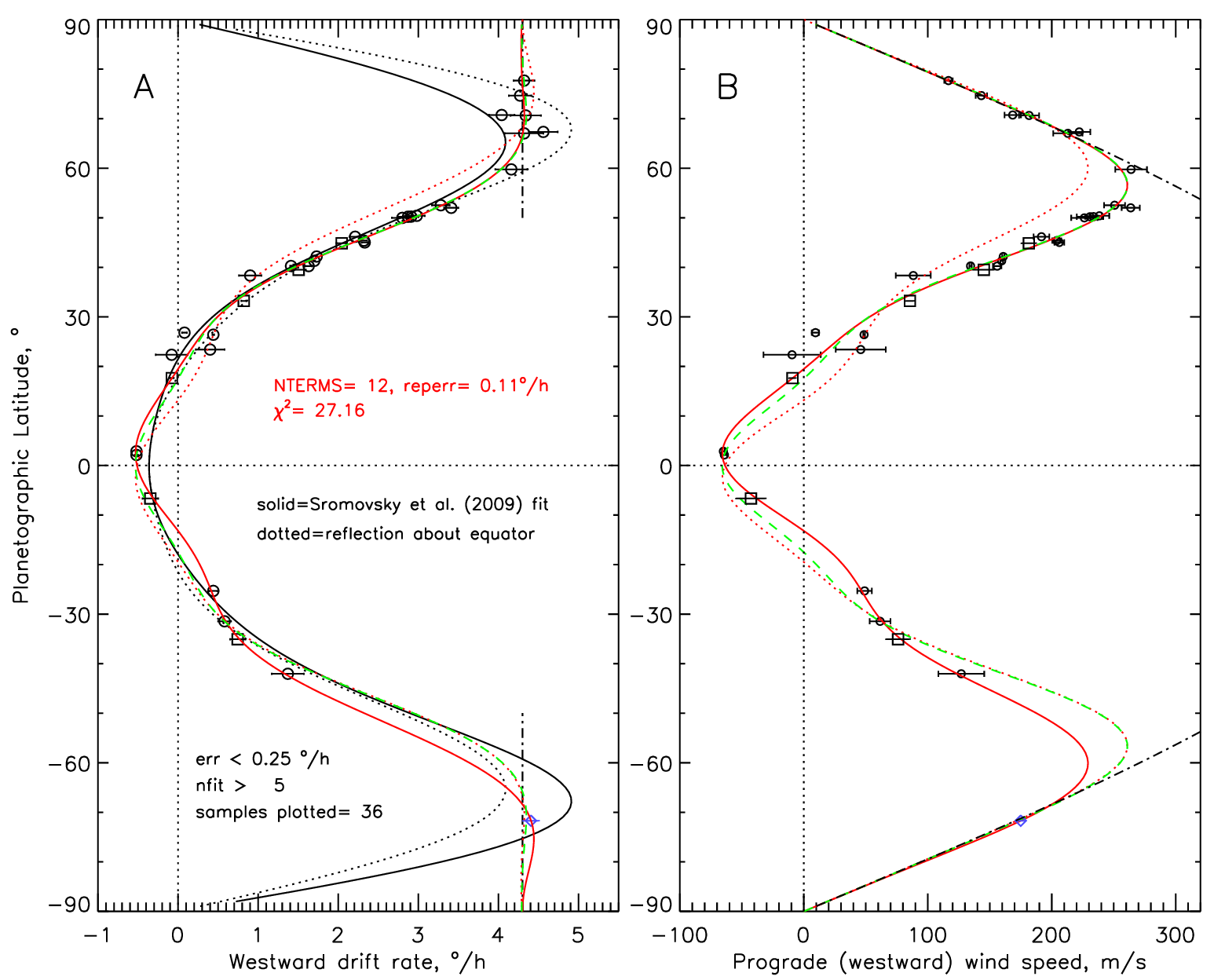}
\caption{A: Combined 2009 HST observations and 2011 Keck and Gemini
  observations of longitudinal drift rates, filtered as indicated in
  the legend, compared to the \cite{Sro2009eqdyn} fit to 2007
  observations (black solid and dotted curves for fit and inverted fit
  respectively), and to a 12-term Legendre polynomial fit to the
  2009-2011 observations (red solid and dotted curves for fit and
  inverted fit respectively) and to an even order (6-term) symmetric fit (dashed
  green curve). The dot-dash lines indicate solid body rotation rates
  of 4.3\degx/h westward relative to the longitude system. The $\chi^2$ value
is for the asymmetric 12-term fit.  See Table\ \ref{Tbl:legfits} for
additional statistical information, which shows that the symmetric
and asymmetric fits are equally good for this case.  B: Corresponding
  observations and fits converted to zonal wind units.}
\label{Fig:legfit9-11}
\end{figure*}

\begin{figure*}[!htb]\centering
\includegraphics[width=6in]{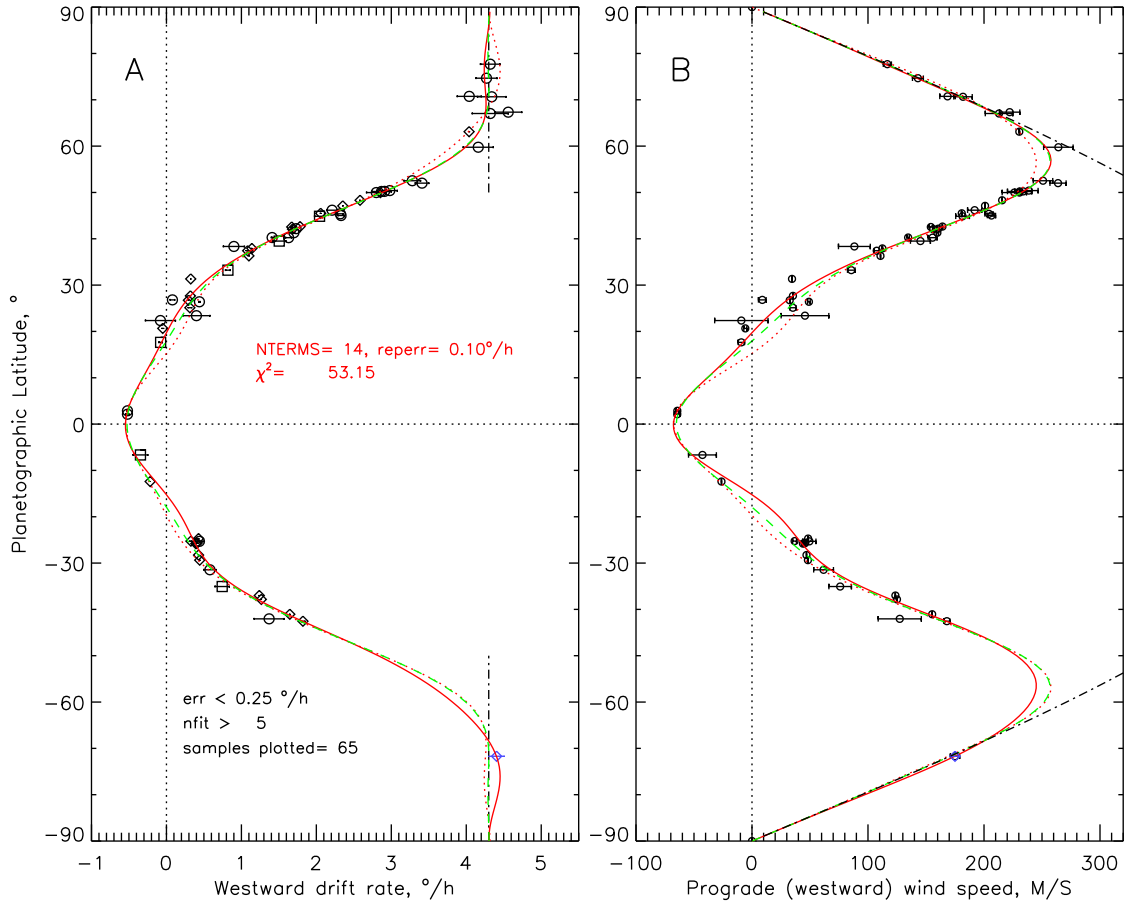}
\caption{As in Fig.\ \ref{Fig:legfit9-11} except that we here
  combined 2009 HST observations and 2007-2011 Keck and Gemini
  observations and carried out a 14-term Legendre polynomial fit to
  the 2007-2011 observations. Here we omitted the \cite{Sro2009eqdyn} fit
for clarity. Again, the symmetric and asymmetric fits are of
indistinguishable quality.}
\label{Fig:legfit7-11}
\end{figure*}

\begin{figure*}[!htb]\centering
\includegraphics[width=6in]{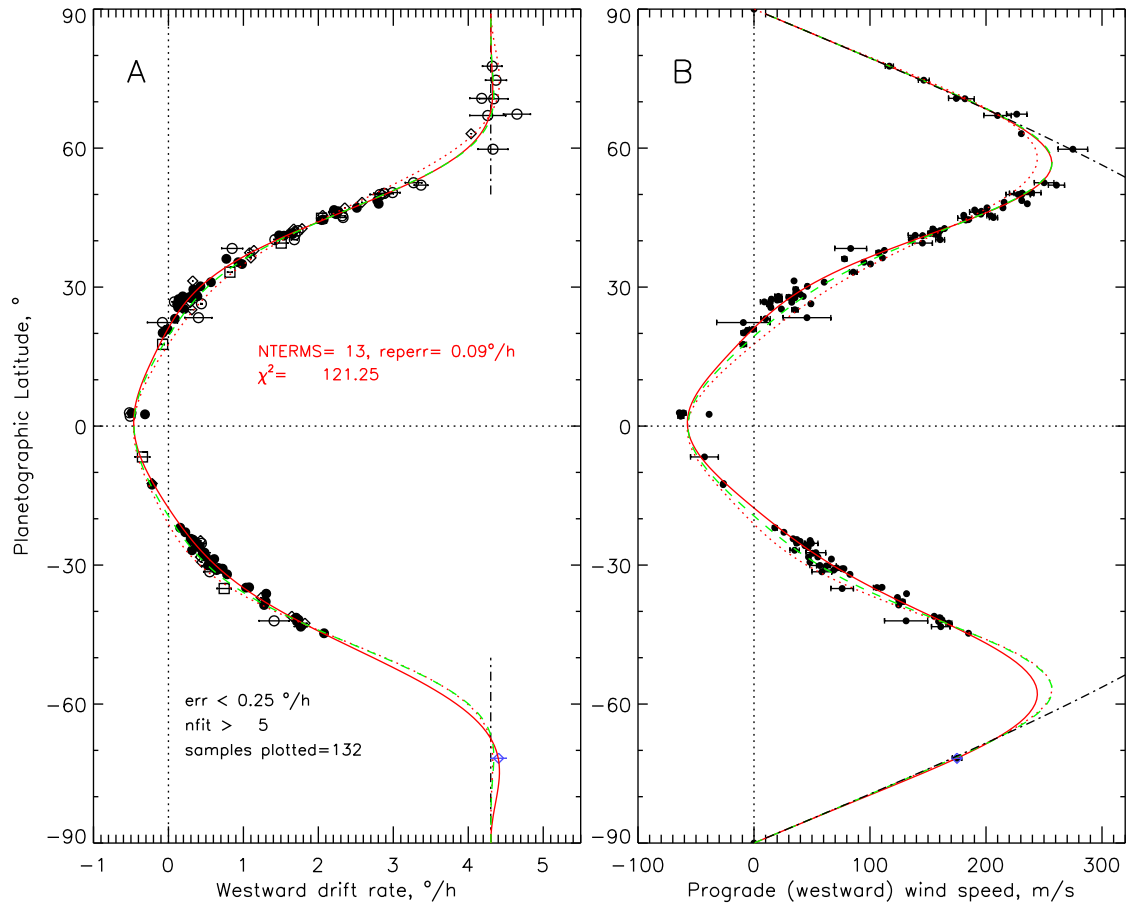}
\caption{As in Fig.\ \ref{Fig:legfit7-11} except that we here
  combined 1997-2005 HST and Keck observations with 2009 HST
  observations and 2007-2011 Keck and Gemini observations and carried
  out a 13-term Legendre polynomial fit.  For this data set,
Table\ \ref{Tbl:legfits} shows that the asymmetric fit is significantly
better than the symmetric fit.}
\label{Fig:legfit86-11}
\end{figure*}

With the constraints described above, we obtained the fits given in
Fig.\ \ref{Fig:legfit9-11} for the 2009 and 2011 data sets combined
and in Fig.\ \ref{Fig:legfit7-11} for the 2007, 2009, and 2011 data
sets combined, with fit coefficients, reperr, and $\chi^2$ values
listed in Table\ \ref{Tbl:legfits}.  The $\chi^2$ uncertainties for
both of these fits are larger than the difference in their $\chi^2$
values, confirming that these data sets cannot distinguish between
symmetric and asymmetric models.  We also combined the data from 1997
through 2005 (see \cite{Sro2009eqdyn} and references therein),
selecting only those with wind errors $<$ 10 m/s and latitude errors
$\le$ 0.5\degx, and then combined these with the high-accuracy winds
from 2007-2011, which yielded a total of 125 observations (127,
including synthetic polar points).  Fits to this combination are shown
in Fig.\ \ref{Fig:legfit86-11}, with coefficients provided in the last
two columns of Table\ \ref{Tbl:legfits}. In this case the asymmetric
fit is far superior to the symmetric fit, with $\chi^2$ being smaller
by 77, which is 3.4 times the expected uncertainty of
$\sqrt{2}\times 16$ in the $\chi^2$ difference.

\begin{table*}\centering
\caption{Legendre polynomial fits to combined high-accuracy drift
  rates from 1997-2011 observations.}
\vspace{0.15in}
\begin{tabular}{c c c c c c c}
\hline
       & \multicolumn{2}{c}{Coefficients (2009-2011)} & \multicolumn{2}{c}{Coefficients (2007-2011)}
                                                  & \multicolumn{2}{c}{Coefficients (1997-2011)}\\
 Order & Even+Odd &  Even only &   Even+odd   & Even only &   Even+odd   & Even only\\
\hline
0 &   1.21686239 &   1.21686239 &   1.25118116 &   1.25328423 &   1.24197012 &   1.25037831\\
 1 &   0.12285565 &              &  -0.02175064 &              &  -0.02848715 &             \\
 2 &   3.59459637 &   0.12285565 &   3.69171880 &   3.71892507 &   3.69457598 &   3.72050211\\
 3 &   0.07697320 &              &   0.07015199 &              &   0.08752786 &             \\
 4 &   0.30820520 &   3.59459637 &   0.13362164 &   0.11135703 &   0.15287708 &   0.12041514\\
 5 &  -0.42303760 &              &  -0.13169439 &              &  -0.13202142 &             \\
 6 &  -0.49582428 &   0.07697320 &  -0.62305001 &  -0.68903134 &  -0.65646542 &  -0.73555624\\
 7 &   0.03484260 &              &  -0.05877913 &              &  -0.08292523 &             \\
 8 &  -0.54870970 &   0.30820520 &  -0.39623823 &  -0.42572871 &  -0.31793598 &  -0.34206461\\
 9 &   0.10603441 &              &   0.01148944 &              &   0.09172810 &             \\
10 &   0.21759050 &  -0.42303760 &   0.30751536 &   0.28876768 &   0.15934681 &   0.20553095\\
11 &   0.07569097 &              &   0.01664328 &              &   0.06504432 &             \\
12 &              &              &  -0.05641039 &   0.04557578 &   0.02752308 &   0.08364616\\
13 &              &              &   0.12180474 &              &              &             \\
NF:     &   27    &   33         &    53        &   60         &  121         &  127\\
reperr & 0.109\degx/h & 0.109\degx/h & 0.105\degx/h & 0.105\degx/h & 0.090\degx/h & 0.088\degx/h\\
$\chi^2$ & 27.16  & 33.90        &  53.15       &  62.92       & 121.34       & 198.065\\
$\chi^2/NF$ &  1.006 &  1.027    &  1.003       & 1.049        & 1.003        &  1.560\\
$\sigma_{\chi^2/NF}$ & 0.27  &  0.25   &  0.19   &  0.18  &  0.13  & 0.13\\[0.05in]
\hline
\\[-0.12in]
\end{tabular}\label{Tbl:legfits}
\parbox{5in}{Note: To limit oscillations in poorly sampled regions, all
  fits make use of a 71\degx S observation by Voyager 2 and fix the
  drift rate at both poles to be 4.3\degx/h westward. 
NF (number of degrees of freedom) is the number of measurements minus
the number of fitted parameters. The representativeness error (reperr)
is discussed in the text.}
\end{table*}

\subsection{Symmetry properties}
 
\begin{figure*}[!htb]\centering
\includegraphics[width=4.5in]{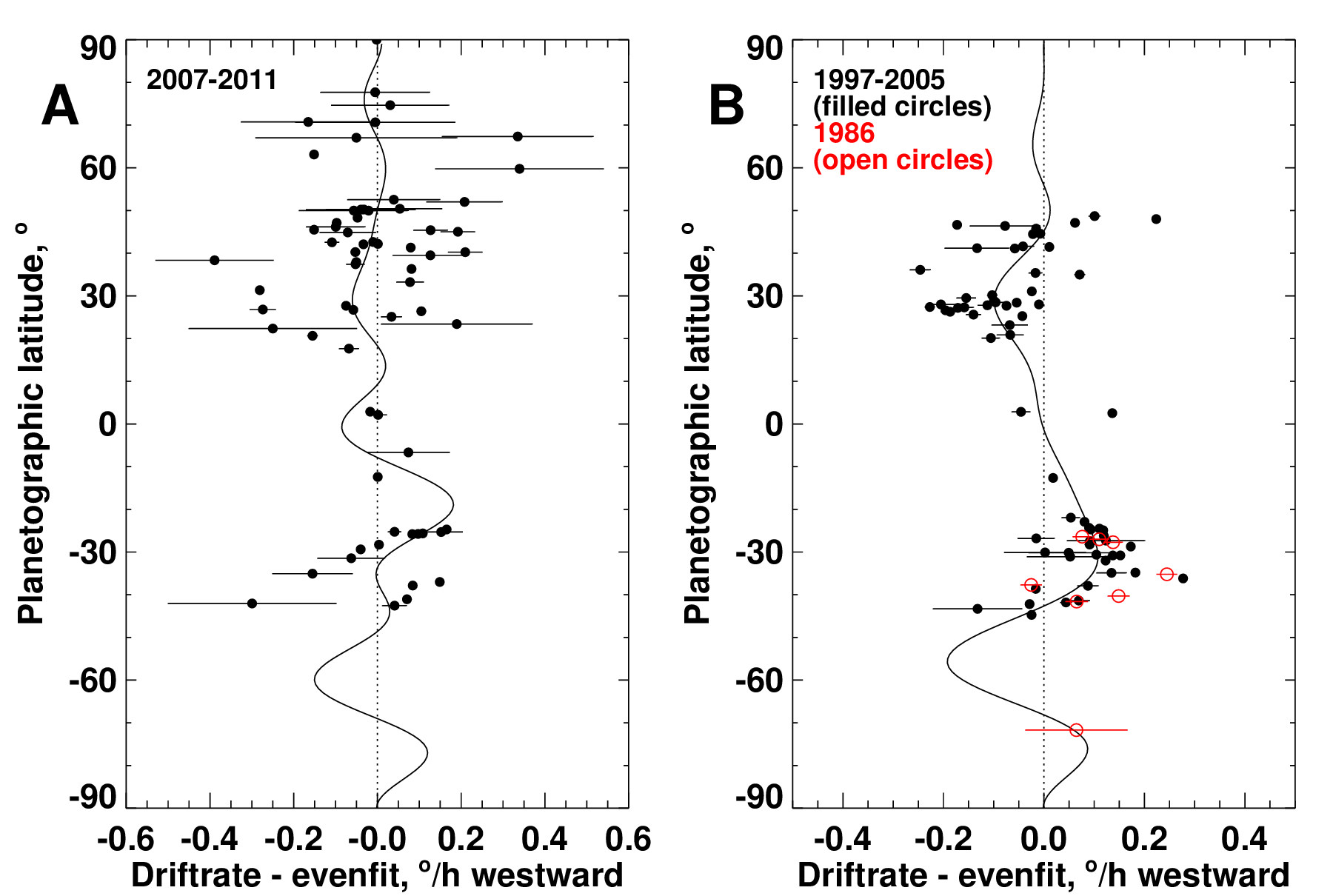}
\caption{A: Using the symmetric fit
of the 2007-2011 observations as a reference and subtracting it from everything
plotted, the non-symmetric fit to those observations appears as the solid curve
and the individual observations from that data set as filled circles.
B: As in A except that the filled circles are from the 1997-2005 data set,
open circles are 1986 Voyager results \cite{SmithBA1986},
and the asymmetric fit is to the largest (1997-2011) data set.}
\label{Fig:asymm}
\end{figure*}

A comparison of the asymmetry properties of the 2007-2011 observations
and those of the earlier observations is provided in
Fig.\ \ref{Fig:asymm}.  The new fits are more symmetric than prior
fits, and the fit that is constrained to be symmetric is very close to
the northern hemisphere fit of the complete fit because there are more
points in the northern hemisphere. However, though the addition of new
observations makes the entire data set somewhat more symmetric, the
remaining asymmetry is better defined. It is not clear whether the
true asymmetry of the zonal winds has changed slightly or whether the
new observations are just sampling a different statistical variation
in target motions.  The better definition of asymmetry obtained from
the larger combined data set depends on the earlier observations,
which are better distributed in latitude and contribute
more measurements in the southern hemisphere.  The large berg feature
is responsible for many of those observations; as it traveled from
33-34\degx S to 8\degx S \citep{Sro2009eqdyn,DePater2011}, it provided
samples of drift rates within that range, though unfortunately sampled
too sparsely to provide a detailed profile.

It is clear that because high-accuracy winds do not always represent
the mean zonal flow with similar accuracy, the asymmetry information
in a sparsely sampled data set can easily be misleading.  With large
numbers of samples, which we obtained by combining all observations
from 1997 onward, this is less of a concern.  By combining
observations between 1997 and 2005 with those up to 2011, we are not
too far from representing the circulation pattern near the 2007
equinox (the mean year of these observations is $\approx$2006).  If
the asymmetry is actually a long-delayed seasonal effect (with a phase
shift near 90\degx), then it should be
reflected about the equator at the prior equinox, and near the
midpoint of the transition in 1986, when Voyager observations were
made. However, as shown by the Voyager results plotted in
Fig.\ \ref{Fig:asymm}B (open circles) it does not appear that the 1986
Voyager measurements are measurably different from our grand average
profile, nor from the 2007 equinox results alone
\citep{Sro2009eqdyn}. This further enhances the probability that the
asymmetry may be a relatively stable feature over a long period,
perhaps more than a uranian year.  While this seems to violate our
expectations of symmetry in the annual average (over the uranian
year), there are other examples of persistent asymmetry in zonal
circulation profiles, namely those of Jupiter and Saturn.  

There are alternative interpretations, however, for the lack of change
in asymmetry of the wind profile.  For example, if the phase shift is closer
to 45\deg and had a sinusoidal variation, then the asymmetry be be the same
at both equinox and solstice, with maxima in between. Another possibility
is that the symmetry variation does not proceed in a uniform or sinusoidal
fashion, but achieves most of the change by the solstice, and changes little
between solstice and the next equinox.  Further observations over many years
will be required to distinguish these possibilities.

\section{Polar cloud features}

\subsection{North polar projections}

\begin{figure*}[!htb]\centering
\includegraphics[width=4.5in]{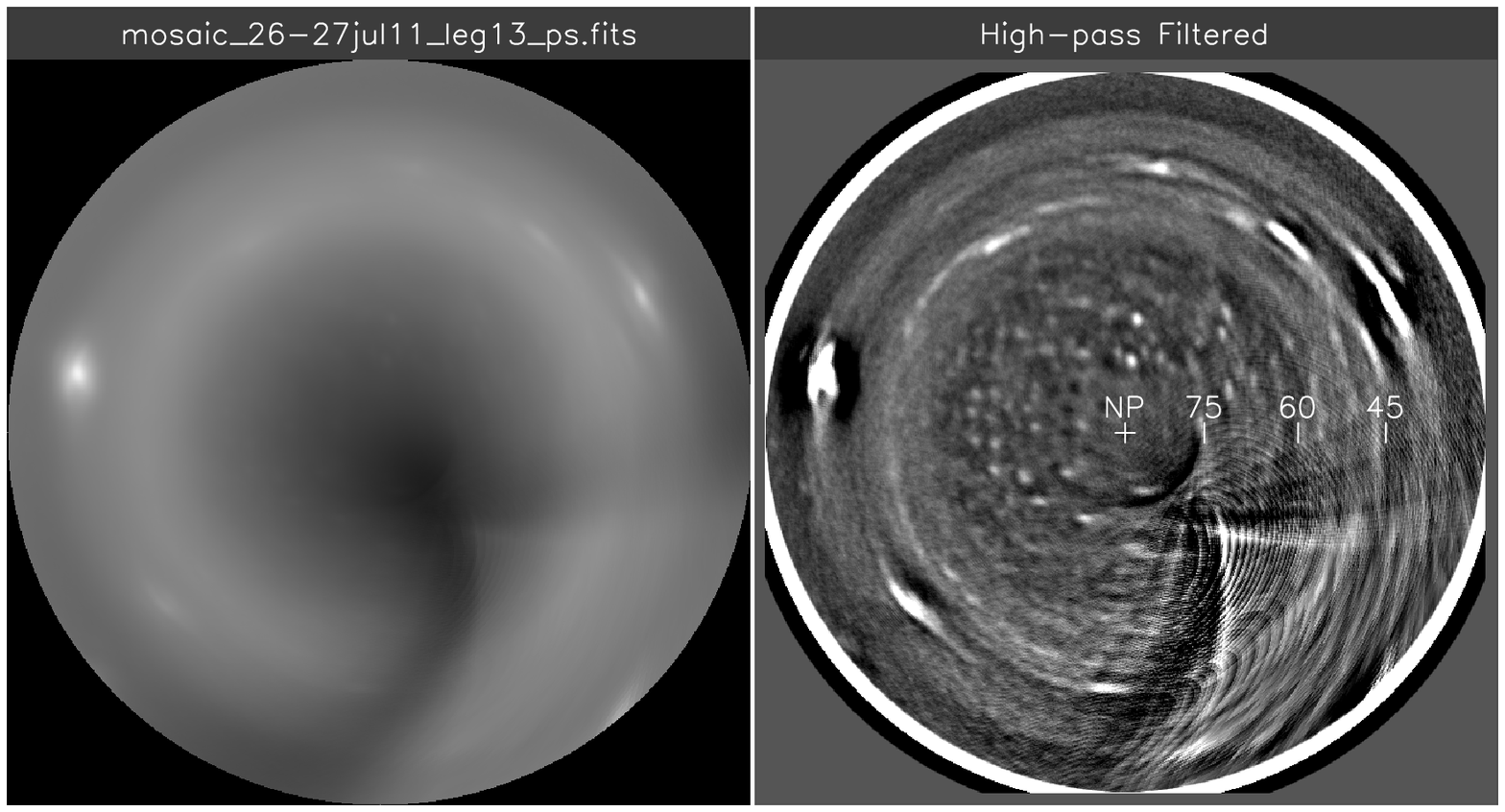}
\caption{Left: polar projection of Keck NIRC2 H-band images of Uranus, blended
  together by shifting longitudes in proportional to the 13-term
  Legendre fit. Right: a high-pass
  filtered version of the left image obtained by subtracting a
  25-pixel boxcar smoothed version from the original image. The
  displayed brightness range for the filtered image is 50 times smaller
  than for the unfiltered image. Both cover the range from 30\degx N
  to 90\degx N.}
\label{Fig:polarcomp1}
\end{figure*}

The 2011 Keck observations provided the best views to date of the north
polar region of Uranus.  This is best illustrated by polar stereographic
projections of the northern hemisphere that we assembled from 85 images
taken on 26 July 2011 and 4 images from 27 July 2011. To combine together
images taken at different times during the rotation of the planet
we had to account for the advection of features by the zonal
wind over the time interval between images.  This is a significant
effect within the north polar jet, where wind speeds exceed 250 m/s
and longitudinal drift rates exceed 5\degx/hour.  In Fig.\ \ref{Fig:polarcomp1}
we display two polar stereographic projections, one with the image brightness proportional
to I/F (left) and the same image with a 25-pixel box car smoothed image
subtracted (right).  In the left hand image latitudinal variations in
brightness are much larger than variations due to local cloud features,
while in the high-pass filtered version, the local cloud features
are revealed in striking detail.  The stretch used in the filtered
image is just -20 to 40 DN, while in the unfiltered image it is 2000-5000 DN,
a factor of 50:1. Thus the small cloud features visible in the filtered
image produce brightness variations that are only a few percent as
large as the latitudinal variations.

\begin{figure*}[!htb]\centering
\includegraphics[width=4.5in]{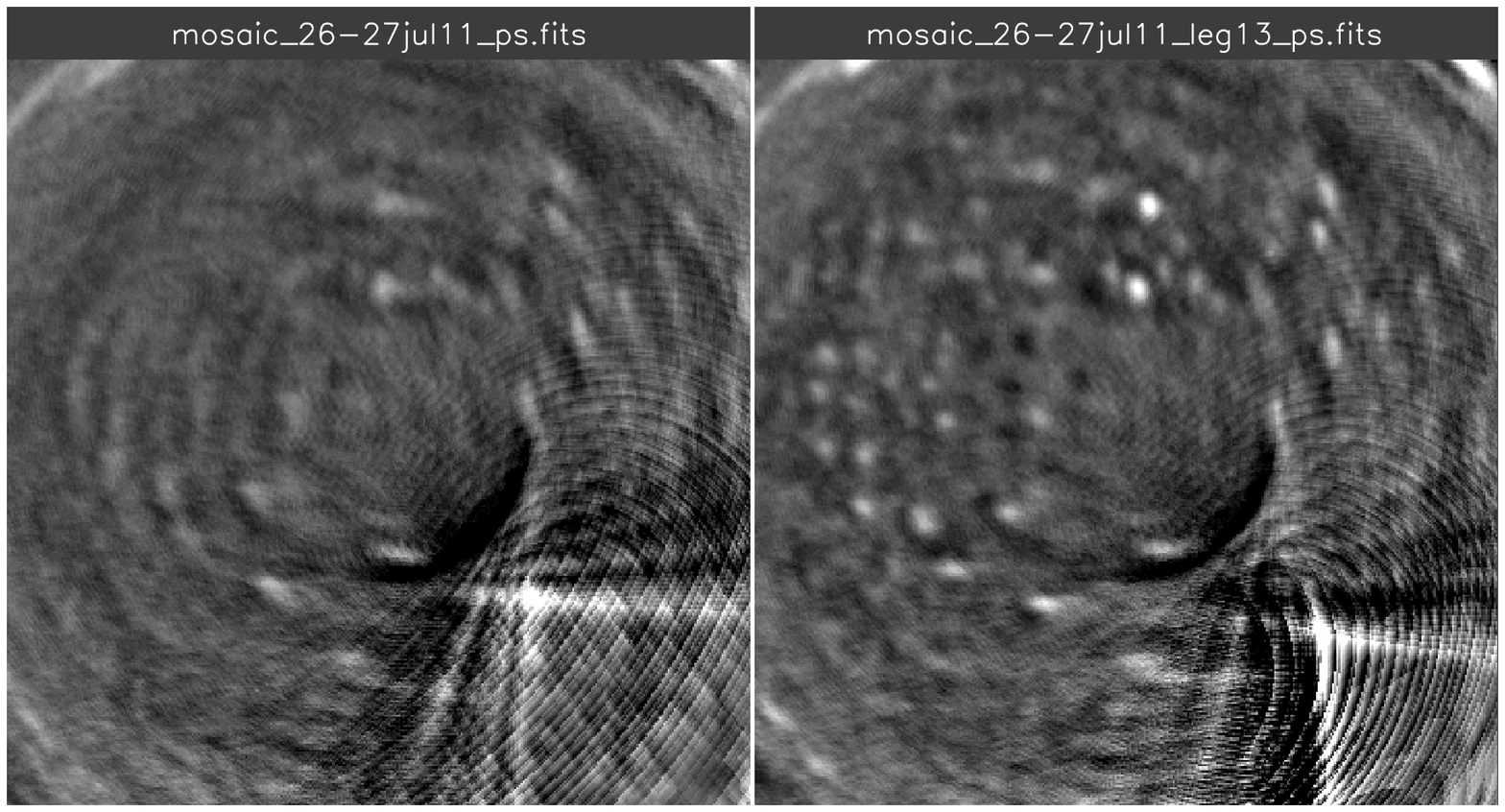}
\caption{Left: Blowup of a
  polar projection mosaic of Keck NIRC2 H-band images of Uranus,
  blended together assuming zero wind speed and high-pass filtered
  with a 25x25 pixel boxcar. Right: X2 blowup of the same polar
  projection mosaic obtained by shifting longitudes in proportion to
  the individual displacements predicted from the 13-term Legendre fit.}
\label{Fig:polarcomp2}
\end{figure*}

The distribution of the small bright features is relatively uniform
over the upper left quadrant of the polar mosaic.  The decline in
numbers of bright features in the other quadrants may simply be a
result of fewer images contributing there.  These are also
regions where the shape of discrete features becomes more distorted by
shear in the wind profile.  The real features don't get sheared, but
the zonal wind shift application causes an artificial distortion that
will reduce contrast. The importance of shifting image pixels to
account for the zonal wind profile is illustrated in
Fig.\ \ref{Fig:polarcomp2}, where the left mosaic was assembled with
no wind shift, while the right was assembled with the nominal wind
shift (the 13-term fit in Table\ \ref{Tbl:legfits} for the 1997-2011
observations).

\begin{figure*}[!htb]\centering
\includegraphics[width=4.5in]{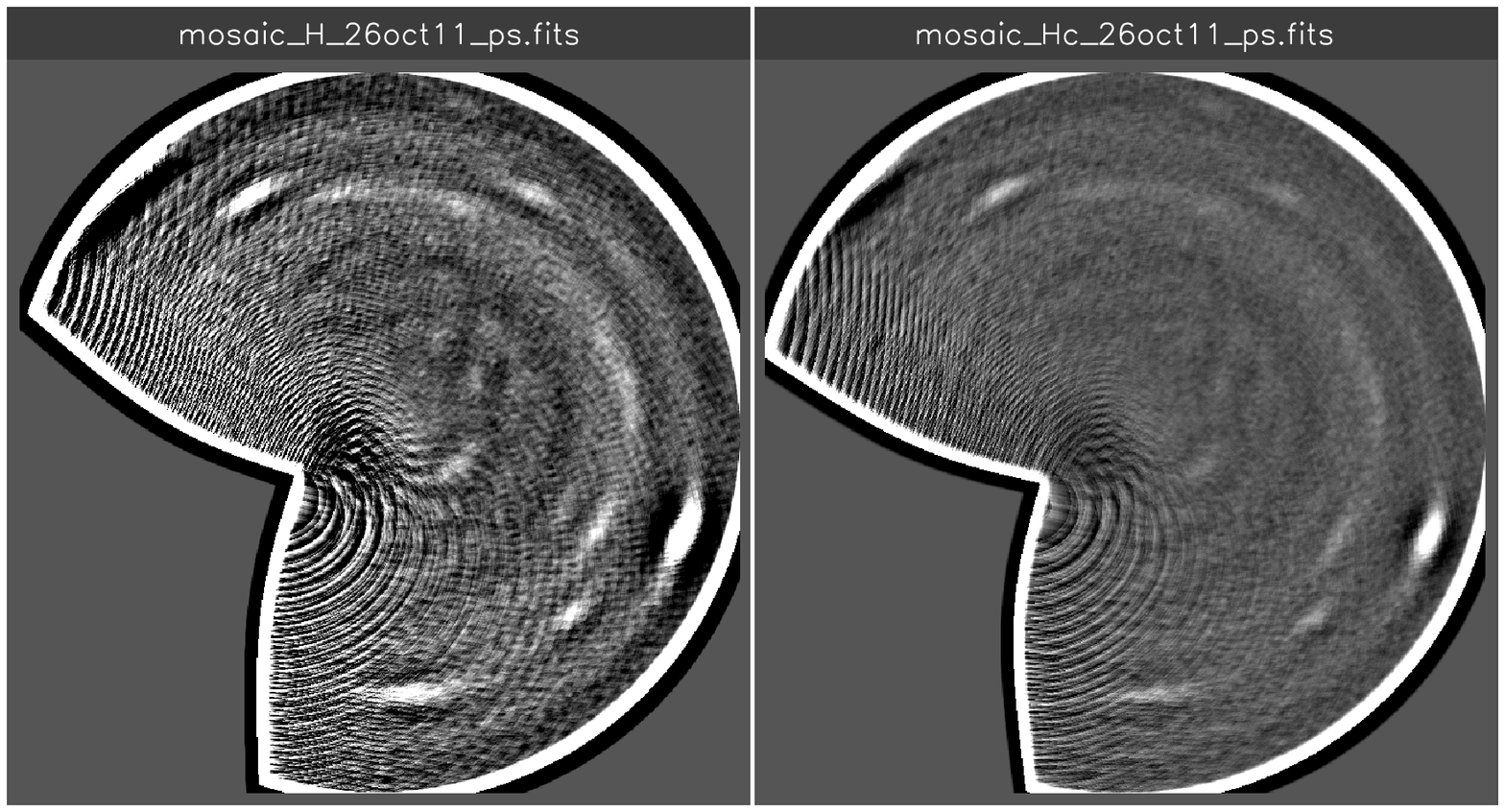}
\caption{Left: Gemini-North H filter polar stereographic projection
from 30\degx N to 90\degx N mosaic average of 33 images. Right: as at left, except
for use of 64 Hcont images.  Both are shown with high-pass filtering obtained
by subtracting a 25-pixel $\times$ 25-pixel boxcar smoothed version from
  each. Each image is 720 pixels $\times$ 720 pixels. }
\label{Fig:gempole}
\end{figure*}

Polar projections of Gemini-North images are displayed in
Fig.\ \ref{Fig:gempole}, which shows a 33-image average for the H
filter and a 64-image combination for the Hcont filter on 26 October
2011.  Both filters show the same discrete feature structure in the
polar regions, but not as fine a detail as displayed by the Keck polar
composite.  Thus, each feature in the Gemini composite is likely a
blurred mix of several features of somewhat smaller scale.  Both filters
display the same pattern of features, proving that they are not due to
random noise fluctuations.  Applying the same technique to Gemini NIRI
images obtained on 25 October revealed no discrete polar features.
Although we had only 19 H images and 18 Hcont images to combine for that day,
the lack of features is more likely due to the lower seeing quality.
On 25 October, the natural seeing (given by the header keyword
AOSEEING) was generally in the 0.5-1 arcsecond range, while on 26
October, the seeing was much more stable and for almost every image
was below 0.4 arcseconds.  Although the 26 July 2011 Keck seeing was
worse, generally in the 0.6-1.2 arcsecond range (at 0.5 \mumx ), the
ability of the Keck AO system to use the planet as a wavefront
reference (VMAG=5.5) instead of a satellite (VMAG=13.7-14.2) allowed
it to produce more detailed images than was possible with the Gemini
AO system under superb natural seeing conditions.  Another virtue of
the Keck AO system is that AO correction deteriorates with angular
distance from the wavefront reference to the target
\citep{DePater2004io}, which is generally larger for a satellite than for
the planet itself.

\subsection{Styles of northern cloud features}

\begin{figure*}[!htb]\centering
\includegraphics[width=4.5in]{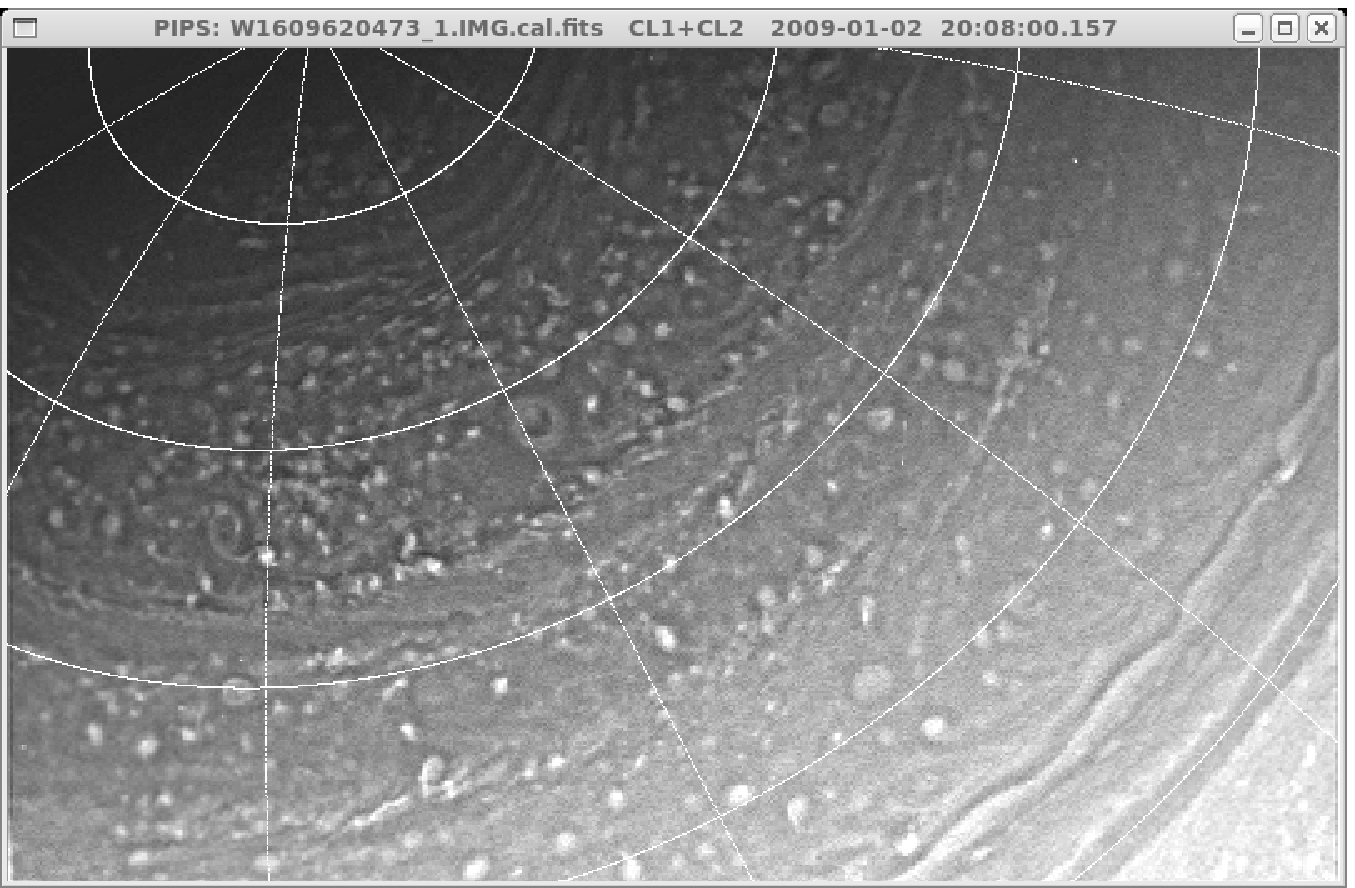}
\caption{Cassini ISS WA image of Saturn's north polar region on 2
  January 2009, when it resembled the North polar region of Uranus.
This image was taken through a clear filter at a distance of 895\,000
  km to Saturn's center and at a phase angle of 59\degx.  The size of
  a single pixel is about 100 km.  Grid spacing is 10\deg in latitude
  and 30\deg in longitude.  
The image is spatially filtered to enhance small scale contrast
 using an unsharp mask in which the raw image is high-pass filtered
 with a box-car smoothing over 13 pixels, then 5 times that filtered
 image is added back to the original image. The large donut feature
 is at 68\degx N, where 1\deg of longitude is 325 km and 1\deg of
latitude is about 950 km. The most numerous small features are only
  about 200 km in diameter.}
\label{Fig:saturn}
\end{figure*}

While the polar projection image is dominated by zonal bands and
zonally streaked features south of about 60\degx N, at higher
latitudes the dominant feature style is small bright spots and a
smaller number of small dark spots. This morphology is similar to what
is seen at high polar latitudes on Saturn \citep{West2009satbook},
where such features were likened to terrestrial cumulus convective
clouds. A Cassini image of Saturn's north polar region, shown in
Fig.\ \ref{Fig:saturn}, bears a striking resemblance to the north
polar region of Uranus, as shown in Fig.\ \ref{Fig:polarcomp2}. On
Saturn, both polar regions can display these features at the same
time, even when the two polar regions are in opposite seasons.  In
January 2009, which was close enough to Saturn's August 2009 equinox
that both polar regions were partially illuminated, Cassini took
images (PIA10583 and PIA10585) revealing small scattered bright spots
in both the spring-time (north) polar region and in the fall (south)
polar region. Uranus' polar regions, on the other hand, are very
asymmetric, as will be shown in the next section.

While the uranian features bear a superficial resemblance to fair
weather cumulus cloud fields on earth, the scale is much larger.  The
spacing between the Uranus polar features is typically 5-10\deg in
longitude and 3-5\deg in latitude. The typical size of the cloud
elements is 1\deg in latitude and 3\deg in longitude. The latitudinal
extent of the features corresponds to a physical distance of $\sim$450
km.  The saturnian features are typically somewhat smaller ($\sim$200
km).  The optical depth of uranian bright cloud features is not well
known, and how much less optical depth is present in the dark regions
is completely unknown.  The fact that none of these features is
visible in K$'$ images, means that they are deeper than the methane
condensation level near 1.2 bars, which is confirmed in a later
section by comparison of H and Hcont images. The few bright polar features
we could analyze do not appear to be optically thick, making it less
likely that they are deep convective features.  On the other hand, it
is possible that our images have not resolved these features, which
would reduce their apparent I/F values and their apparent optical
depths.  The saturnian features appear very opaque in 5-\mum images,
implying that they are optically thick.

\subsection{Comparison of north and south polar regions}

\begin{figure*}[!htb]\centering
\includegraphics[width=4.5in]{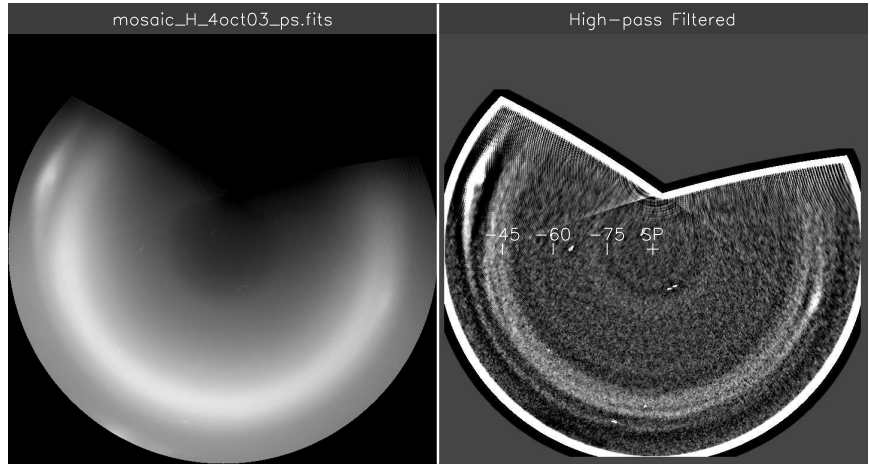}
\caption{Left: South polar projection of Keck NIRC2 H-band images of
  Uranus obtained on 4 October 2003, blended together by shifting
  longitudes in proportional to the 10-term Legendre fit wind profile
  of \cite{Sro2009eqdyn}. Right: a high-pass filtered version of the
  left image obtained by subtracting a 25-pixel boxcar smoothed
  version from the original image, which is 720 pixels on a side. The
  displayed brightness range for the filtered image is 30 times smaller
  than for the unfiltered image. Both cover the range from 30\degx S
  to 90\degx S.}
\label{Fig:southpole}
\end{figure*}

Although post-Voyager imaging has failed to reveal any discrete clouds
in the south polar region of Uranus, there remained a small possibility that
subtle features were present but not seen because of
insufficient S/N.  To address that issue we gathered high quality
images from 2003, and formed a south polar mosaic using the same wind
shift approach used for the north polar mosaic.  While we didn't have
as many images to work with, the quality of individual images was
generally higher because the AO system was better adjusted and the
seeing may have been better.  The result for mosaicking the three H
images from 4 October is shown in Fig.\ \ref{Fig:southpole}.  In the
overlap region where all three images contribute to reduce noise
levels (roughly a 100\deg region centered at 7 o'clock), there is no
indication of any discrete cloud feature standing above the noise
level.  This should be compared to the north polar mosaic displayed in
Fig.\ \ref{Fig:polarcomp1}.  There is certainly no feature in the
south polar region with the contrast found in the north polar region
in 2011.  Instead, the south polar region seems devoid of discrete
cloud features, but has (or at least had in 2003) a banded structure
that is currently not seen in the north polar region.  A south polar
mosaic created without accounting for zonal wind shift produced an
image that is virtually identical to the one produced with the wind
shift.  Neither showed any small discrete features.  Even better views
of the south polar region, provided by HST images from 1994-2000
\cite{Rages2004}, also failed to reveal any discrete polar cloud
features.  

Although the observed polar asymmetry on Uranus might be unrelated to
seasonal forcing, as perhaps is the case for the wind asymmetry, a
seasonal cause for the polar asymmetry is certainly the most
plausible.  If it is a seasonal effect, then a few years after the
next equinox, when the south polar region will again come into view,
we would expect it to look like the north polar region looks now, and
just before that equinox we would expect the north polar region to be
devoid of discrete cloud features.  As we continue to observe the
north polar region for the next four decades, at some point the
convective activity should stop, and a polar cloud cap should form
\citep{Hammel2007var}. Exactly when these transitions will take place
and what detailed mechanisms are involved in this expected response to
seasonal forcing remain to be determined.

The dramatically different appearance of the two polar regions of
Uranus is a possible indication of different stability structures of
the atmospheres.  \cite{Kark2009IcarusSTIS} reported
reduced methane mixing ratio at high southern latitudes, based on 2002
STIS spectra.  This finding was confirmed by \cite{Sro2011occult}, who
also found the depletion to be relatively shallow (only down to a few
bars).  These results suggested a meridional flow of gas from low
latitudes upward through the methane condensation level, where
condensation reduces the methane mixing ratio, transport of that
depleted gas to high latitudes, and then descent of the depleted gas,
which reduces the average methane mixing ratio to the observed
levels. Such a circulation would tend to suppress convection in the
region of down-welling gas, which is consistent with the absence of any
discrete convective features in the south polar region in 2003.
Although this is suggestive, the real reason for lack of convection in
the south might be due to increased thermal stability produced by some
other mechanism that alters the thermal structure in this region, which
itself might have a seasonal origin.  It is worth noting that there
are occasions when convective storms can occur in regions of 
general downwelling. On Jupiter, for example, thunderstorms occur
primarily in belts \citep{Showman2005ammonia}.

The meridional flow suggested above for the upper troposphere, is
opposite to the direction of heat transport required to reduce the
temperature difference between equator and pole.  According to
\cite{Friedson1987}, the average annual solar energy absorbed by the
poles exceeds that absorbed at the equator, and radiative convective
models suggest that the poles should be $\sim$6 K warmer than the
equator, which significantly exceeds the $2.5$ K obtained with
meridional transports included.  If northern poleward meridional
transport of methane is occurring in the upper troposphere and
southward transport is occurring somewhat deeper, that suggests that
most of the heat transport is occurring in the return flow.  But
deeper still, another poleward transport is suggested by microwave
observations down to the 50-bar level \citep{Hofstadter2004DPS}, where
microwave absorbing gases (NH$_3$ and H$_2$S) seem to be symmetrically
depleted in both polar regions.

\section{Vertical structure}

\subsection{The spatial modulation method}

An estimate of the pressure levels of many of the discrete cloud
features can be made using the ratio of spatial modulations in H and
Hcont filters, as employed in the simplest form by \cite{DePater2011}
and in a somewhat more exact form by \cite{Sro2012bs}.  These two
filters not only have different penetration depth profiles
(Fig.\ \ref{Fig:penprof}) over a useful pressure range, but they also
have sufficiently similar effective wavelengths that we can ignore
wavelength dependent differences in seeing or effective spatial
resolution, which can distort the observed modulation ratios and the
inferred cloud pressures.  The similarity in spatial resolution is
only valid if the seeing is stable over the time span covered by the H
and Hcont images.

Fig.\ \ref{Fig:penprof} displays the I/F obtained when a unit-albedo
surface is placed at different depths within a model aerosol-free
Uranus atmosphere.  For a given filter, the apparent (externally
viewed) I/F of a unit albedo surface at pressure P approaches its
asymptotic value as P becomes arbitrarily large (which yields the same
I/F as for a clear atmosphere) at a rate that depends on the degree of
absorption within the filter band.  The I/F difference from the
asymptotic value indicates how sensitive the filter is to spatial
modulations in cloud reflectivity or amount as a function of where
they occur in pressure. In the simplest case of a small isolated cloud
feature, the modulation can be thought of as simply the difference
between the peak I/F of a cloud feature and the I/F of its
surroundings.  The useful pressure range for H and Hcont is between
300 mb and 4-5 bars, between which the I/F differences in H decline
more rapidly with pressure than the I/F differences in Hcont. The
ratio of H modulations to Hcont modulation thus provides a measure of
the pressure location of the modulations.

In applying the spatial modulation technique to our observations, we
followed \cite{Sro2012bs} in calculating model ratios for middle
latitudes for an array of cloud pressures and view angles including
the effects of the main cloud layer between 1.2 and 2 bars
\citep{Sro2011occult}.  The discrete cloud is treated as a physically
thin perturbation inserted into the background structure.  Both the
background and discrete perturbations are modeled as conservative Mie
particle layers of 0.8 \mum in radius with a refractive index of 1.3.
An effective pressure for the discrete cloud can be inferred from the
ratio of spatial modulations in I/F (see Fig. 15 of Sromovsky et al.
2012).  From the peak-to-peak amplitude in modulations we can infer the change in
optical depth required.  If the cloud feature is not spatially
resolved, the inferred optical depth maximum will be less than
the true optical depth.

\subsection{Cloud structure results}

\begin{figure}[!htb]\centering
\includegraphics[width=3.4in]{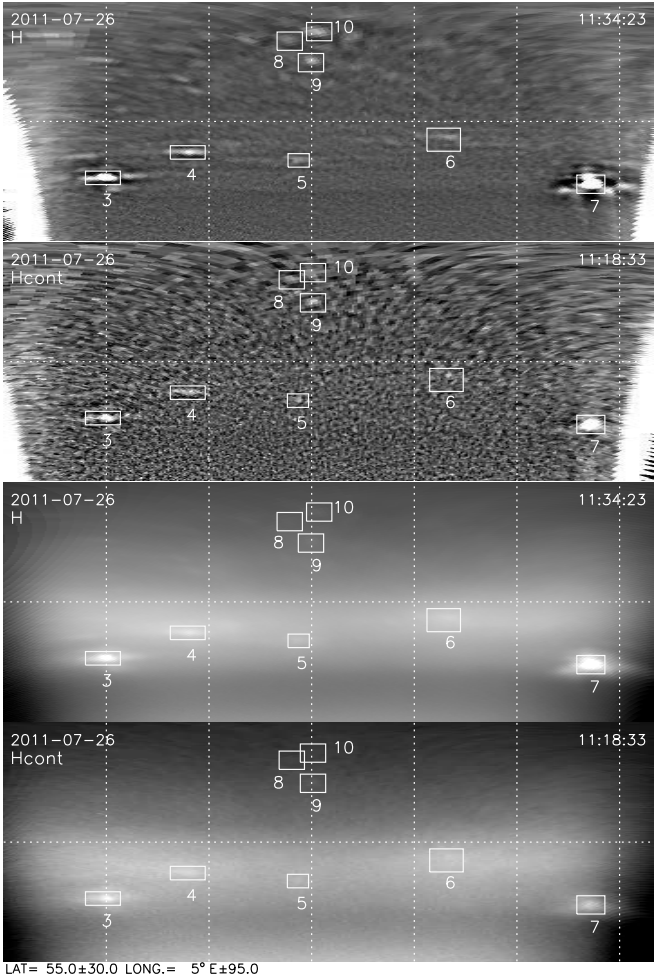}
\caption{Rectilinear projection of NIRC2 NA images of northern cloud
  features on 26 July 2011, covering longitudes from -90\degx E to
  100\degx E and latitudes from 25\degx N to 85\degx N, with labeled
  outlines indicating target boxes for analyzing spatial
  modulations. ID numbers refer to rows in
  Table\ \ref{Tbl:cloudpars}. The top two images are high-pass
  filtered versions of the bottom images, which display local discrete
  cloud variations more clearly.  Note that the S/N ratio of the Hcont
  image limits the number of high latitude features for which a
  pressure estimate can be obtained.}
\label{Fig:spacemodim}
\end{figure}

\begin{figure*}[!htb]\centering
\includegraphics[width=4.5in]{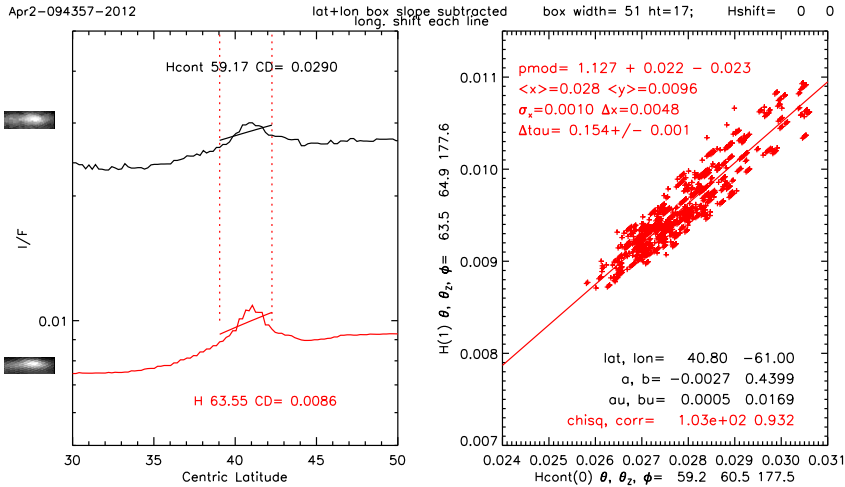}
\caption{I/F vs latitude (left) for H and Hcont through target box 3
  (in Fig.\ \ref{Fig:spacemodim}) and correlation plot of I/F
  observations within that target box (right), with linear regression
  fit line and inferred pressure (pmod) in bars and optical depth
  ($\Delta$tau).}
\label{Fig:spacemodcorr}
\end{figure*}

A sample application of the modulation ratio technique to target 3 in
Table\ \ref{Tbl:cloudpars} is provided in Figs.\ \ref{Fig:spacemodim}
and \ref{Fig:spacemodcorr}.  A target box size is selected and
positioned in the Hcont image (shown twice in
Fig.\ \ref{Fig:spacemodim}, once without and once with high-pass
filtering).  The corresponding target box in the H image is shifted
slightly as needed to maximize correlation between Hcont and H
variations.  This shift is usually no more than a fraction of a degree
of longitude and less than a few tenths of a degree latitude.  Once
the correlation is maximized, we fit a straight line to the H vs Hcont
I/F variations within the target box to determine the slope (see
Fig.\ \ref{Fig:spacemodcorr}) at the view angles of the observations.
We then use a spline interpolation of the model slope vs. pressure and
view angles to infer a pressure from the observed slope.  From the
pressure we interpolate the models of I/F vs optical depth to infer
the optical depth change that would yield the observed variation of
I/F at the observed view angles.  For target 3 in
Fig.\ \ref{Fig:spacemodim} the effective pressure is 1.17$\pm$0.02
bars (formal error), and the optical depth variation is 0.15, with
negligible formal uncertainty.  Note that I/F gradients defined by the
boundary of the target box are subtracted from the gradients within
the box to keep background variations from affecting local changes.

The inferred pressure of modulation is affected by the relative
calibration error between the H and Hcont images.  Our analysis
measures the central disc value of each image and then converts the
image to I/F units by a scale factor that produces a central disk I/F
equal to previously determined I/F values.
New central disc
measurements using preliminary calibrations of 2011 Gemini-North
images, indicate significant changes from those of \cite{Sro2007struc}.
The new values are 0.86$\times10^{-2}$
(H) and 2.9$\times10^{-2}$ (Hcont) with relative errors of 5\% and
10\% respectively.  
The effect of a 10\% increase in the ratio of H to Hcont on derived
pressures in a typical example is to decrease P by about 100 mb and to
increase the derived optical depth by 30\%.  Thus the formal
uncertainties given in Table\ \ref{Tbl:cloudpars} are generally less
significant than the uncertainties associated with relative
calibrations, which add about 0.1 bars to the pressure uncertainty and
30\% to the optical depth uncertainty. There are also bias errors
associated with unresolved cloud features, for which apparent optical
depths will be lower than actual values.

\begin{figure}[!htb]\centering
\includegraphics[width=3.2in]{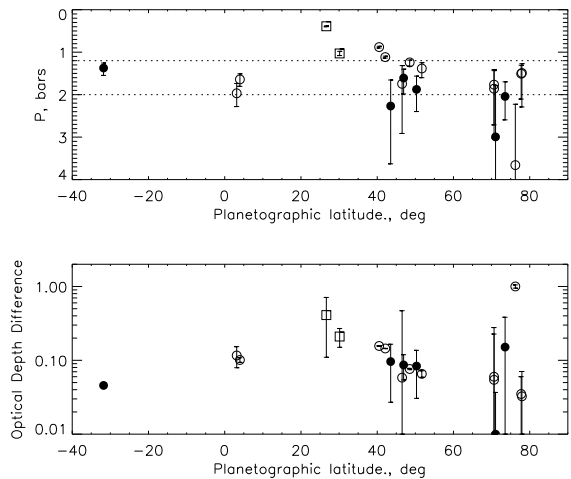}
\caption{Discrete cloud pressure (top) and optical depth perturbation (bottom)
inferred from the ratio and amplitude of spatial modulations in H and Hcont
images. Open circles are from Keck NIRC2 images taken on 26 July 2011,
filled circles from Gemini NIRI images taken on 26 October 2011, and
squares are from \cite{Sro2012bs}.}
\label{Fig:cloudpars}
\end{figure}

The results in Table\ \ref{Tbl:cloudpars} are also plotted in
Fig.\ \ref{Fig:cloudpars}, where we see that most clouds reside deeper
than the 1.2-bar methane condensation level, including the broad
equatorial features and all the compact polar cloud features for which
we could estimate the pressure.  Bright spots at northern middle
latitude are found to be well above the methane condensation level and
thus likely contain a component of methane ice.  Very few of the cloud
features have significant optical depths.  Even the brightest feature
(BS1) has less than one optical depth.  A more detailed study of the
high-latitude features will require improvements in the S/N of the
Hcont images.

\begin{table}\centering
\caption{Summary of discrete cloud parameters for 26 July 2011 (1-12) and 26 October 2011 (A-F).}
\begin{tabular}{c c c c c c c}
\hline\\[-0.1in]
PID &TID & PClat & PGlat & Elon & P, bars & Optical depth\\[0.01in]     
\hline\\[-0.1in]
   1& & 3.80  &  3.98& -6.2&  1.64$^{+0.16}_{-0.13}$ &  0.101$\pm$0.01\\[0.05in]
   2& & 3.00  &  3.14& 39.0&  1.97$^{+0.31}_{-0.23}$ &  0.116$\pm$0.04\\[0.05in]
   3&21 &40.8 & 42.12&-61.0&  1.12$^{+0.02}_{-0.02}$ &  0.145$\pm$0.001\\[0.05in]
   4&20 &47.2 & 48.52&-36.2&  1.24$^{+0.09}_{-0.06}$ &  0.076$\pm$0.002\\[0.05in]
   5& 5 & 45.2& 46.53& -4.0&  1.75$^{+1.2}_{-0.4}$ &  0.058$\pm$0.4\\[0.05in]
   6& 18& 50.4& 51.70& 39.4&  1.38$^{+0.22}_{-0.14}$ &  0.066$\pm$0.01\\[0.05in]
   7& 13& 39.2& 40.51& 81.6&  0.88$^{+0.02}_{-0.02}$ &  0.156$\pm$0.001\\[0.05in]
   8& & 75.6  & 76.23& -5.8&  3.66$^{+6.3}_{-1.4}$ &  1.000$\pm$0.05\\[0.05in]
   9& & 69.8  & 70.65&  0.4&  1.77$^{+0.95}_{-0.36}$ &  0.054$\pm$0.2\\[0.05in]
  10& & 77.4  & 77.95&  0.4&  1.49$^{+0.80}_{-0.22}$ &  0.033$\pm$0.04\\[0.05in]
  11& & 77.2  & 77.76&  0.0&  1.51$^{+0.60}_{-0.20}$ &  0.035$\pm$0.025\\[0.05in]
  12& & 69.8  & 70.65&  0.2&  1.86$^{+1.2}_{-0.4}$ &  0.060$\pm$0.2\\[0.05in]
   A& & 45.6  & 46.93&227.8&  1.61$^{+0.38}_{-0.21}$ &  0.087$\pm$0.03\\[0.05in]
   B& & 49.0  & 50.31&246.8&  1.88$^{+0.52}_{-0.31}$ &  0.084$\pm$0.05\\[0.05in]
   C& & 42.2  & 43.53&328.0&  2.27$^{+1.4}_{-0.6}$ &  0.096$\pm$0.07\\[0.05in]
   D& & 72.8  & 73.54&252.6&  2.04$^{+0.55}_{-0.34}$ &  0.151$\pm$0.2\\[0.05in]
   E& & 70.2  & 71.03&297.2&  2.99$^{+7.0}_{-1.2}$ &  0.010$\pm$0.03\\[0.05in]
   F& &-30.6  &-31.78&286.0&  1.37$^{+0.17}_{-0.12}$ &  0.046$\pm$0.003\\[0.05in]
 BS1& & 25.6  & 26.65&129.2&  0.39$^{+0.01}_{-0.01}$ &  0.410$\pm$0.3\\[0.05in]
 BS2& & 29.0  & 30.14&207.0&  1.03$^{+0.05}_{-0.05}$ &  0.210$\pm$0.06\\[0.05in]
\hline
\hline
\end{tabular}\label{Tbl:cloudpars}\vspace{0.1in}
\parbox{3.2in}{NOTES: BS1 and BS2 parameters are from \cite{Sro2012bs}
  and represent the brightest parts of the features. PID is the ID for
  pressure estimation. TID is the tracking ID where applicable. Other
  headings are as given in Table\ \ref{Tbl:keckwinds}.}
\end{table}

\section{Summary and Conclusions}

Following the 2007 equinox, new circulation measurements were obtained
from imaging observations by HST in 2009, by Keck in July 2011, and by
Gemini-North in October 2011.  The new imaging programs used repeated
short exposures followed by averaging in latitude-longitude
projections to increase signal to noise ratios so that low contrast
cloud features could be detected and tracked. Application of high S/N
imaging and analysis techniques \citep{Fry2012} to 2011 observations
of Uranus using Gemini and Keck telescopes resulted in detection of
many trackable cloud features even under less than ideal observing
conditions (for Keck) and less than ideal AO references (for
Gemini). The 2011 observations, which provide better views of Uranus'
north polar region, have enabled more accurate measurements of the
zonal winds at high latitudes.  Using 2011 Keck images in July and
Gemini images in October, we made measurements of cloud motions using
a variety of methods, using manual tracking and guided correlation
tracking, as well as using high-pass filtered images to enhance small
discrete features and median differenced images to enhance larger
discrete features, which often appear at low latitudes.  Our main
conclusions are as follows:

\begin{enumerate}

\item Zonal winds measured in 2009 and 2011 are roughly consistent
  with the 10-term Legendre polynomial fit to 2007 observations
  \citep{Sro2009eqdyn}, but deviate in a direction that provides
  somewhat better agreement with the profile reflected about the
  equator, which is in the direction of reducing hemispheric asymmetry, although
this is not yet statistically significant.

\item The best high latitude results were obtained from Keck 2011 imagery,
and clearly define a prograde jet peak near 260 m/s at a planetographic latitude
near 60\degx N.

\item Cloud motions in the north polar region, from the jet peak
  northward, appear most consistent with solid body rotation at a rate
  of 4.3\degx/h westward relative to the interior.  The Voyager wind
  measurement at 71\degx S (the only south polar measurement) is
  consistent with the same rotation period.

\item Using manually guided correlation measurements, we obtained several high
  accuracy measurements of near equatorial (between 2 and 3\deg N)
  drift rates averaging 0.52$\pm$0.02\degx/h eastward in 2011; this
  corresponds to a retrograde wind speed of 64$\pm$2 m/s, which is in
  close agreement with new measurements in 2003 Keck
  images obtained by \cite{Hammel2005winds}.

\item The near-equatorial features (2-4\deg N) are broad and diffuse,
  and form a pattern consistent with a wavenumber of 9 (40\deg spacing
  in longitude), suggesting that the measured motions might differ
  from the mean mass flow.  The 2011 pattern is not complete,
  however, and thus the \cite{Hammel2005winds} wavenumber 12 pattern
  (also incomplete but better constrained) might be consistent with
  our results.

\item When 2011 observations are combined with 2009 HST wind results
  of \cite{Fry2012}, the resulting fits provide less asymmetry at high
  latitudes than found by \cite{Sro2009eqdyn}, mainly because of
  better constraints obtained from high-latitude measurements.  For
  this data set, the difference in fit quality between symmetric and
  asymmetric fits is statistically insignificant.

\item When combined with past observations from 1997 onward, which
  provides better sampling at southern mid latitudes, we find a small
  but well defined asymmetry at middle latitudes, extending from about
  15\deg to 40\degx.  Longitudinal drift rates at 35\deg N differ from
  drift rates at 35\deg S by an average of 0.2\degx/h eastward, which
  corresponds to a wind speed difference of 20 m/s.  There is no
  substantial evidence regarding symmetry properties at high latitudes, primarily due
  to  a lack of samples in the southern hemisphere.

\item Voyager 1986 wind measurements do not differ significantly from
  the zonal wind profile established by 1997-2011 observations,
  raising the possibility that the asymmetry at the most recent
  equinox is a long-term feature of Uranus' circulation, or that the
  asymmetry has a phase shift relative to seasonal forcing that is
  less than 90\degx, or that the asymmetry response function is itself
  asymmetric, with perhaps most of the asymmetry reversal occurring
  early in the seasonal cycle. 
These possibilities can be distinguished by measurements between now
and the next solstice in 2030.

\item High accuracy wind measurements at nearby latitudes often differ by far more
than their estimated uncertainties, which might be a result of different
atmospheric depths, latitude differences between the generating circulation feature
and the observed cloud feature, and possible long-period eddy motions.

\item The morphology of cloud features in the north polar region in
  2011 is very different from the south polar region observed in 2003.
  Although the south polar region was devoid of discrete cloud
  features, the north polar region (north of 60\degx N) contains small
  bright (but low contrast cloud features), widely distributed and
  reminiscent of fair-weather cumulus cloud fields on earth, though on a much
  larger scale, with features approximately 450 km in size, and spaced
  roughly 1500-3000 km apart. This morphology, which resembles
  Saturn's polar regions, is distinct from lower latitudes, where the
  predominant morphology consists of zonal bands with longitudinally
  stretched streaky cloud features, and not widely distributed.

\item The north polar region was also found to have small dark spots comparable in size
to the small bright features, but considerably less numerous.

\item Where S/N ratios permitted measurement we determined that high
latitude bright cloud features were located at pressures at 1.5 bars or
deeper, with the most accurate estimates between 1.4 and 2 bars, which
is within the main background cloud layer observed on Uranus.

\item Cloud pressures inferred from spatial modulations in H and Hcont
images near the equator were between 1.5 and 2 bars, as were many
clouds at northern mid latitudes. 

\item A few of the brighter cloud features were found to be close
to or above the methane condensation level, the highest being
the BS1 feature described by \cite{Sro2012bs}, which reached
an estimated pressure near 400 mb.

\end{enumerate}

The cloud pressure and optical depth estimates could be improved in several
ways.  The first is to compute model ratios for background models that
vary in latitude as needed to reproduce the background I/F, instead of
using the mid-latitude calculations. The second, which applies to the
existing observations, is to determine a more accurate relative
measurement of the central disk I/F of Uranus in H and Hcont spectral
filters, then apply that calibration to the existing
observations.  A third improvement could be obtained by improving the
S/N ratio of the Hcont measurement, which will require new measurements.

The high S/N methods applied to Keck imaging observations will likely
yield even better results with better seeing conditions and a well
adjusted AO system.  According to the test case of \cite{Fry2012}, there is
a great potential for further significant improvements in coverage of the zonal
wind profile.

\section*{Acknowledgments}

We thank Eric Karkoschka and an anonymous reviewer for detailed and
constructive suggestions for improving the paper.  LAS and PMF
acknowledge support from NASA Planetary Astronomy Grant NNX08A051G and
Space Telescope Science Institute Grants HST-GO-11639.04-A and
HST-GO-12463.05-A. HBH acknowledges support by Grants from NASA's
Planetary Astronomy and Atmospheres Programs. IdP acknowledges support
from NASA Grant NNX07AK70G. KAR acknowledges support by grants from
the Space Telescope Science Institute.  This research was partly based
on Hubble Space Telescope observations.  We thank staff at the
W. M. Keck Observatory, which is made possible by the generous
financial support of the W. M. Keck Foundation.  We thank those of
Hawaiian ancestry on whose sacred mountain we are privileged to be
guests. Without their generous hospitality none of our groundbased
observations would have been possible.


\end{document}